\documentclass[journal]{IEEEtran}

\ifCLASSINFOpdf
\else
   \usepackage[dvips]{graphicx}
\fi
\usepackage{url}

\usepackage{graphicx}
\usepackage{algorithm}
\usepackage{stmaryrd}
\usepackage{amssymb,amsmath,amsfonts,bm}
\usepackage{mathrsfs}
\usepackage{booktabs}
\usepackage{xcolor}
\usepackage{multirow}
\usepackage{array,cite}
\usepackage{microtype}
\newcommand{\ie}{\textit{i.e.}}
\newcommand{\eg}{\textit{e.g.}}

\newcommand{\Tr}{\mathsf{T}}
\newcommand{\Hr}{\mathsf{H}}

\newcommand{\ComplexGaussian}[1]
{\mathcal{N}_{\mathbb{C}}\!\left({#1}\right)}

\newcommand{\ComplexStable} [2]
{\mathcal{S}^{#1}_{\mathbb{C}}\!\left({#2}\right)}

\newcommand{\ComplexStudent}[2]
{\mathcal{T}^{#1}_{\mathbb{C}}\!\left({#2}\right)}
\newcommand{\ComplexGG}[2]
{\mathcal{GG}^{#1}_{\mathbb{C}}\!\left({#2}\right)}

\newcommand{\ComplexGH}[2]
{\mathcal{GH}^{#1}_{\mathbb{C}}\!\left({#2}\right)}

\newcommand{\InverseGamma}[2]{\mathcal{IG}\!\left({#1}, {#2}\right)}

\newcommand\GeneralizedInverseGaussian[3]{\mathcal{GIG}\!\left({#1},{#2},{#3}\right)}

\newcommand{\mbG}{\mathbf{G}}
\newcommand{\mbH}{\mathbf{H}}
\newcommand{\mbI}{\mathbf{I}}
\newcommand{\mbJ}{\mathbf{J}}

\newcommand{\mbQ}{\mathbf{Q}}

\newcommand{\mbV}{\mathbf{V}}
\newcommand{\mbW}{\mathbf{W}}
\newcommand{\mbX}{\mathbf{X}}
\newcommand{\mbY}{\mathbf{Y}}

\newcommand{\mba}{\mathbf{a}}

\newcommand{\mbe}{\mathbf{e}}

\newcommand{\mbg}{\mathbf{g}}

\newcommand{\mbq}{\mathbf{q}}

\newcommand{\mbu}{\mathbf{u}}
\newcommand{\mbv}{\mathbf{v}}

\newcommand{\mbx}{\mathbf{x}}

\newcommand{\mbz}{\mathbf{z}}

\newcommand{\mbmu}{\boldsymbol{\mu}}

\newcommand{\mbTh}{\boldsymbol{\Theta}}

\newcommand{\mbSi}{\boldsymbol{\Sigma}}

\newcommand{\mbPh}{\boldsymbol{\Phi}}
\newcommand{\mbPs}{\boldsymbol{\Psi}}

\begin{document}


\title{
Generalized Fast Multichannel Nonnegative Matrix Factorization
Based on Gaussian Scale Mixtures\\
for Blind Source Separation
}

\author{Mathieu Fontaine, \IEEEmembership{Member, IEEE}, Kouhei Sekiguchi, \IEEEmembership{Member, IEEE}, Aditya Arie Nugraha, \IEEEmembership{Member, IEEE}, Yoshiaki Bando, \IEEEmembership{Member, IEEE}, Kazuyoshi Yoshii, \IEEEmembership{Member, IEEE}
\thanks{
    Manuscript received XXX YYY, 2022; revised XXX YYY, 2022; accepted XXX YYY, 2022.
    Date of publication XXX YYY, 2022; date of current version XXX YYY, 2022.
    This work was partially supported by
    	JSPS KAKENHI Nos.~19H04137, 20K19833, and 20H01159,
	    and NII CRIS Collaborative Research Program operated by NII CRIS and LINE Corporation.
    The associate editor coordinating the review of this manuscript and approving it for publication was Prof. xxx.
    \textit{(Corresponding author: Mathieu Fontaine.)}}
\thanks{
	Mathieu Fontaine is with LTCI, Télécom Paris, Institut Polytechnique de Paris, France and with the Center for Advanced Intelligence Project (AIP), RIKEN, Tokyo 103-0027, Japan
	(e-mail: mathieu.fontaine@telecom-paris.fr).}
	
\thanks{Kouhei Sekiguchi and Aditya Arie Nugraha are with the Center for Advanced Intelligence Project (AIP), RIKEN, Tokyo 103-0027, Japan
(e-mail: kouhei.sekiguchi@riken.jp; adityaarie.nugraha@riken.jp)}	
\thanks{
	Kazuyoshi Yoshii is with the Center for Advanced Intelligence Project (AIP), RIKEN, Tokyo 103-0027, Japan, and
	the Graduate School of Informatics, Kyoto University, Kyoto 606-8501, Japan
	(e-mail: yoshii@i.kyoto-u.ac.jp).}
\thanks{
	Yoshiaki Bando is with the National Institute of Advanced Industrial Science and Technology (AIST), Tokyo, 135-0064, Japan
	(e-mail: y.bando@aist.go.jp).}
}

\markboth{IEEE/ACM Transactions on Audio, Speech, and Language Processing, Vol. xx, No. xxx, xxxx 2022}
{Shell \MakeLowercase{\textit{et al.}}: Bare Demo of IEEEtran.cls for IEEE Journals}
\maketitle

\fussy
\allowdisplaybreaks[4]

\begin{abstract}
This paper describes heavy-tailed extensions 
 of a state-of-the-art versatile blind source separation method
 called fast multichannel nonnegative matrix factorization (FastMNMF)
 from a unified point of view.
The common way of deriving such an extension
 is to replace the multivariate complex Gaussian distribution
 in the likelihood function
 with its heavy-tailed generalization, 
 \eg, the multivariate complex Student's $t$ 
 and leptokurtic generalized Gaussian distributions,
 and tailor-make the corresponding parameter optimization algorithm.
Using a wider class of heavy-tailed distributions
 called a Gaussian scale mixture (GSM),
 \ie, a mixture of Gaussian distributions 
 whose variances are perturbed by positive random scalars called impulse variables,
 we propose GSM-FastMNMF
 and develop an expectation-maximization algorithm 
 that works even when the probability density function of 
 the impulse variables have no analytical expressions. 
We show that existing heavy-tailed FastMNMF extensions
 are instances of GSM-FastMNMF
 and derive a new instance
 based on the generalized hyperbolic distribution that include the normal-inverse Gaussian, Student's $t$, and Gaussian distributions as the special cases.
Our experiments show that 
 the normal-inverse Gaussian FastMNMF outperforms the state-of-the-art FastMNMF extensions and ILRMA model
 in speech enhancement and separation
 in terms of the signal-to-distortion ratio. 
 
\end{abstract}


\IEEEpeerreviewmaketitle

\section{Introduction}

\begin{figure}
\centering
\includegraphics[width=0.85\columnwidth]{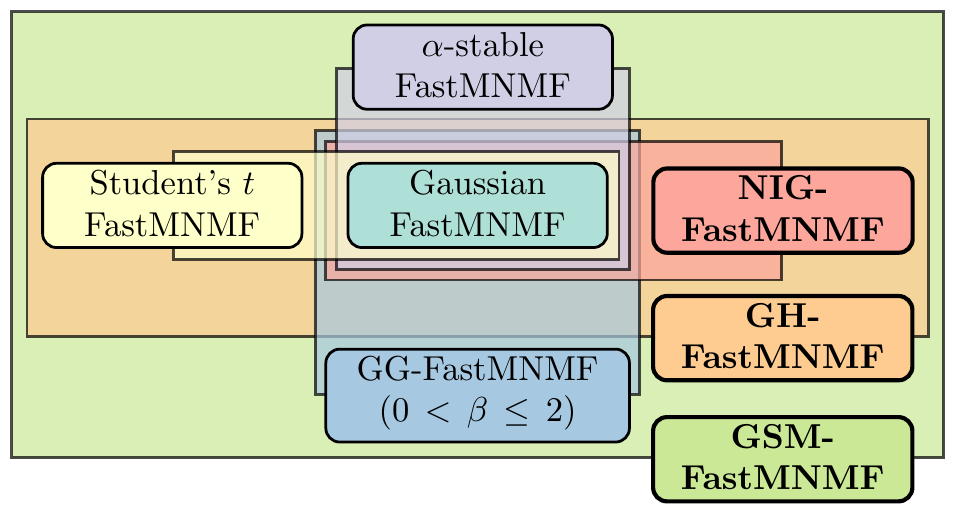}
\vspace{-1mm}
\caption{
Heavy-tailed extensions of FastMNMF. 
We propose a general form
 based on the Gaussian scale mixture representation (GSM-FastMNMF) 
 that includes existing variants
 based on the Student's $t$ distribution ($t$-FastMNMF),
 the $\alpha$-stable distribution ($\alpha$-FastMNMF),
 and the leptokurtic generalized Gaussian (GG) distribution 
 (GG-FastMNMF).
We instantiate a new variant 
 based on the generalized hyperbolic (GH) distribution (GH-FastMNMF)
 and its special case 
 based on the normal-inverse Gaussian (NIG) distribution (NIG-FastMNMF).
}
\label{fig:GSM-FastMNMF}
\end{figure}

The goal of blind source separation (BSS)
 is to estimate latent sources
 from observed mixtures recorded by multiple microphones \cite{vincent18}.
In general, the audio signal is converted to
 a time-frequency (TF) spectrogram
 obtained with short-time Fourier transform (STFT).
A vast majority of modern statistical BSS methods
 are based on the local Gaussian model (LGM)
 that assumes the STFT coefficients of each TF bin to follow 
 a zero-mean multivariate complex Gaussian distribution
 whose covariance matrix is given 
 by the product of the nonnegative \textit{power spectral density} (PSD)
 and the positive semidefinite \textit{spatial covariance matrix} (SCM),
 where the SCM is a full-rank matrix 
 under echoic conditions~\cite{vincent18}.
 
A typical approach to BSS
 is to perform maximum-likelihood (ML) estimation 
 based on a unified probabilistic model of observed mixtures
 consisting of source and spatial models 
 representing the PSDs and SCMs of sources, respectively~\cite{duong2010under}.
Assuming the low-rankness of source PSDs
 as is often the case in real sounds (\eg, music),
 the source model has often been formulated 
 as a LGM with nonnegative matrix factorization (NMF),
 resulting in a versatile BSS method called
 multichannel NMF (MNMF)
 \cite{ozerov2009multichannel,Sawada2013multichannel}. 
One way of reducing the computational cost of MNMF
 stemming from a large number of SCM inversions
 is to restrict the SCMs of all sources to rank-1 matrices,
 resulting in independent low-rank matrix analysis (ILRMA) \cite{kitamura2016determined}.
Another promising way
 is to restrict the source SCMs
 to jointly-diagonalizable yet full-rank matrices~\cite{Ito2018fastfca,Ito2018fastfcaas, sekiguchi2020fast,ito2021joint},
 \ie, to represent the SCM of each source
 as a conical sum of common rank-1 SCMs,
 resulting in FastMNMF
 \cite{sekiguchi2020fast,ito2021joint}.
Although FastMNMF (denoted as $\mathcal{N}$-FastMNMF) 
 outperforms ILRMA under echoic conditions,
 the light-tailed LGM inherited from MNMF
 does not fit impulsive sounds with a large dynamic range. 

To improve the robustness of $\mathcal{N}$-FastMNMF 
 against such perturbations,
 local heavy-tailed models have often been used instead of the LGM
 \cite{kitamura2016student,
 kamo2020jointstudent,kitamura2018generalized,
 kamo2020jointsubgaussian,mogami2019independent,
 leglaive2018student, fontaine2018multichannel,
 fontaine2020unsupervised,fontaine2021alpha} (Fig.~\ref{fig:GSM-FastMNMF}).
Using a local Student's $t$, leptokurtic generalized Gaussian (GG), 
 or $\alpha$-stable model, 
 $\mathcal{N}$-FastMNMF~\cite{sekiguchi2020fast,ito2021joint}
 can be extended to 
 $t$-FastMNMF~\cite{kamo2020jointstudent}, 
 leptokurtic GG-FastMNMF~\cite{kamo2020jointsubgaussian}\footnote{
The generalized Gaussian (GG) distribution with a shape parameter $\beta > 0$
 consists of leptokurtic and platykurtic (heavy- and light-tailed) sub-families.
In \cite{kamo2020jointsubgaussian},
 only platykurtic GG-FastMNMF with $\beta \in [2, 4)$ is described,
 but leptokurtic GG-FastMNMF with $\beta \in (0, 2]$ can also be derived straightforwardly
 in the same way that 
 their generalized gaussian ILRMA extensions
 are derived from ILRMA~\cite{mogami2019independent}
 (Section~\ref{sec:GG-FastMNMF}).
 },
 or $\alpha$-FastMNMF~\cite{fontaine2020unsupervised,fontaine2021alpha}\footnote{
 The original version of $\alpha$-FastMNMF~\cite{fontaine2020unsupervised}
 considers the source-specific time-frequency-varying impulsiveness,
 whereas its modified version~\cite{fontaine2021alpha}
 considers the source-specific time-varying but frequency-invariant impulsiveness.
To explain heavy-tailed extensions of FastMNMF from a unified point of view, 
 in this paper we discuss another version of $\alpha$-FastMNMF
 that considers the source-independent time-frequency-varying impulsiveness (Section~\ref{sec:alpha-FastMNMF}).
 }, respectively.
Similarly, 
 the LGM in ILRMA~\cite{kitamura2016determined}
 can be replaced by a Student $t$~\cite{kitamura2018generalized},
 leptokurtic GG~\cite{mogami2019independent},
 and $\alpha$-stable~\cite{fontaine2021alpha} local model, respectively\footnote{
 ILRMA~\cite{kitamura2016determined} 
 is exactly a special case of FastMNMF~\cite{sekiguchi2020fast},
 whereas the Student $t$~\cite{kitamura2018generalized}
 and leptokurtic GG~\cite{mogami2019independent} extensions
 based on the product of \textit{univariate} heavy-tailed distributions for \textit{independent} sources
 are not special cases of their respective heavy-tailed FastMNMF model in~\cite{kamo2020jointstudent}
 and \cite{kamo2020jointsubgaussian}
 based on \textit{multivariate} heavy-tailed distributions
 for \textit{dependent} sources.
 In \cite{fontaine2021alpha},
 a rank-1 version of FastMNMF with an $\alpha$ model is naively qualified as ILRMA extension,
 but the sources are only \textit{conditionally} independent.
In this paper
 we take this approach to deriving a rank-1 version of X-FastMNMF
 and call it X-R1-FastMNMF (\eg, $t$-R1-FastMNMF instead of $t$-ILRMA for Student's $t$ model) to avoid confusion.
 }.
For ML estimation with $t$- and GG-FastMNMF, 
 deterministic parameter optimization algorithms with closed-form update rules
 have been tailor-made
 according to the minorization-maximization (MM) principle.
Note that all the Student's $t$, leptokurtic GG, and $\alpha$-stable distributions 
 belong to the \textit{Gaussian scale mixture} (GSM) family~\cite{andrews1974scale};
 a random vector following a GSM
 can be represented as a Gaussian random vector 
 whose scale is perturbed by a positive random variable
 called an \textit{impulse variable}~\cite{csimcsekli2015alpha}. 
For ML estimation with $\alpha$-FastMNMF, in contrast,
 the compound GSM representation is used
 for addressing the non-closed-form probability density function (PDF) 
 of the $\alpha$-stable distribution~\cite{fontaine2020unsupervised},
 but calls for a stochastic Metropolis-Hastings (MH) step
 for optimizing the impulse variables~\cite{robert2013monte}.
Note that the GSM model has been studied 
 for audio source separation \cite{fevotte2007bayesian}, 
 speech enhancement \cite{hao2009speech},
 and sparse signal representation \cite{godsill2007bayesian},
 but not within the FastMNMF framework.


In this paper,
 we propose a general form of heavy-tailed FastMNMF 
 based on the GSM representation (GSM-FastMNMF) 
 that encompasses the aforementioned heavy-tailed FastMNMF extensions
 and a new heavy-tailed variant
 based on the generalized hyperbolic (GH) distribution~\cite{barndorff1977infinite, barndorff1982normal} (GH-FastMNMF).
A noticeable instance of GH-FastMNMF
 is one based on the normal-inverse Gaussian (NIG) distribution (NIG-FastMNMF),
 which was experimentally proven to perform best for speech enhancement and separation.
 Recent studies in \cite{dhull2021expectation} and \cite{ghasami2020autoregressive} for instance make use of NIG and GH innovations respectively within an autoregressive model for time series modeling. 
 The ML estimation is done through an expectation-maximization (EM) framework as in \cite{karlis2002type,oigaard2005estimation} for NIG and \cite{protassov2004based, palmer2016algorithm} for GH model respectively. 

For ML estimation with GSM-FastMNMF,
 we propose a general parameter optimization algorithm
 based on the EM principle and called multiplicative update variational expectation-maximization (MU-VEM).

This readily instantiates a closed-form parameter estimation algorithm
 for the above-mentioned variants except for $\alpha$-FastMNMF,
 which have been tailor-made independently.
The key advantage of this technique
 is that closed-form update rules might be obtained
 even when the impulse variable law is unknown or analytically intractable. 
 
The rest of the paper is organized as follows.
Section \ref{sec:FastMNMF} reviews 
 existing variants of FastMNMF. 
Section \ref{sec:GSM-FastMNMF} formulates GSM-FastMNMF
 and instantiates GH-FastMNMF and NIG-FastMNMF.
Section \ref{sec:evaluation} 
 compares the existing and new variants of FastMNMF
 in speech enhancement and speaker separation.
Section \ref{sec:conclusion} concludes the paper 
 while a short Appendix provides PDFs and proofs of mathematical results used in this article.

\section{Existing Variants of Fast Multichannel\\Nonnegative Matrix Factorization}
\label{sec:FastMNMF}

We review a versatile BSS method called MNMF~\cite{Sawada2013multichannel}
 that maximizes the multivariate complex Gaussian likelihood (denoted by $\mathcal{N}_{\mathbb{C}}$)
 and its computationally-efficient special case called FastMNMF~\cite{sekiguchi2020fast}.
We also introduce heavy-tailed extensions of FastMNMF
 that maximize the multivariate complex Student's $t$, 
 leptokurtic GG, and $\alpha$-stable likelihoods
 (denoted by $\mathcal{T}_{\mathbb{C}}^{\nu}$, $\mathcal{GG}_{\mathbb{C}}^{\beta}$, 
 and $\mathcal{S}_{\mathbb{C}}^{\alpha}$, respectively).



\subsection{Problem Specification} 

Suppose that $N$ sources are recorded by $M$ microphones. 
Let $\mbX_n \triangleq \{\mbx_{nft}\}_{f,t=1}^{F,T} \in \mathbb{C}^{F \times T \times M}$ 
 be the multichannel complex spectrogram of source $n \in \{1,\dots,N\}$ (called a source \textit{image}),
 where $\triangleq$ represents equality by definition,  $F$ and $T$ represent the number of frequency bins and that of time frames, respectively.
Let $\mbX \triangleq \{\mbx_{ft}\}_{f,t=1}^{F,T} \in \mathbb{C}^{F \times T \times M}$
 be that of the observed mixture.
Assuming the additivity of sources 
 in the STFT domain,
 our goal is to estimate the source images $\{\mbX_n\}_{n=1}^N$ from the mixture $\mbX$ such that
\begin{align}
 \mbx_{ft} = \sum_{n=1}^{N}\mbx_{nft}.
 \label{eq:mix_image}
\end{align}

\subsection{Probabilistic Formulation} 
\label{sec:problem_specification}

The standard approach to BSS is 
 based on the local Gaussian model (LGM)~\cite{duong2010under}. 
Assuming both the independence of sources and that of time-frequency bins,
 the source image $\mbx_{nft} \in \mathbb{C}^M$ of source $n$ at frequency $f$ and time $t$
 is assumed 
 to independently follow a zero-mean multivariate circularly-symmetric complex Gaussian distribution as follows (see Eq~\eqref{eq:pdf-N} for the PDF):
\begin{align}
 \mbx_{nft} \sim 
 \ComplexGaussian{\lambda_{nft} \mbG_{nf} \triangleq \mbY_{nft}},
 \label{eq:p_x_nft}
\end{align}
where $\ComplexGaussian{\mbmu, \mbSi}$ denotes the multivariate complex Gaussian distribution
 with a mean vector $\mbmu$
 and a covariance matrix $\mbSi \succeq 0$
 ($\bm\mu$ is omitted for brevity if $\bm\mu = \bm{0}$)
,
 $\lambda_{nft} \ge 0$ 
 is the \textit{power spectral density} (PSD) of the source $n$
 at frequency $f$ and time $t$ denoted $s_{nft}$,
 and $\mbG_{nf} \succeq \bm{0}$ is the positive semidefinite 
 \textit{spatial covariance matrix} (SCM) 
 of source $n$ at frequency $f$.
Note that $\succeq$ stands for the set of positive semidefinite matrices.
Let $\bm\Lambda \triangleq \{\lambda_{nft}\}_{n,f,t=1}^{N,F,T}$ 
 and $\mbG \triangleq \{\mbG_{nf}\}_{n,f=1}^{N,F}$
 be the sets of the source PSDs and SCMs, respectively.
 
Using the law stability by linear combination of independent Gaussian vectors,
 Eqs.~\eqref{eq:mix_image} and \eqref{eq:p_x_nft} give
 the mixture $\mbx_{ft}$ distributed as follows:
\begin{align}
 \mbx_{ft} 
 \sim 
 \ComplexGaussian{\sum_{n=1}^{N} \lambda_{nft} \mbG_{nf} \triangleq \mbY_{ft}},
 \label{eq:p_x_ft}
\end{align}
where $\lambda_{nft}$ and $\mbG_{nf}$ are represented
 by \textit{source} and \textit{spatial} models, respectively,
 as described in Section \ref{sec:source_and_spatoal_models}.
Given the mixture $\mbX$ as observed data,
 we aim to estimate $\bm\Lambda$ and $\mbG$
 that maximize the likelihood for $\mbX$ given by Eq.~\eqref{eq:p_x_ft}.


BSS is implemented with a Wiener filter 
 that computes the posterior distribution of $\mbx_{nft}$ given $\mbx_{ft}$
 as follows:
\begin{align}
 &\mbx_{nft} \mid \mbx_{ft}
 \nonumber\\
 &\sim 
 \ComplexGaussian{\mbY_{nft} \mbY_{ft}^{-1} \mbx_{ft}, \mbY_{nft} - \mbY_{nft} \mbY_{ft}^{-1} \mbY_{nft}}.
\end{align}
The maximum-a-posteriori (MAP) estimate of the source image $\mbx_{nft}$
 is thus given by 
 $\mathbb{E}[\mbx_{nft} \mid \mbx_{ft}] = \mbY_{nft} \mbY_{ft}^{-1} \mbx_{ft}$.


\subsection{Source and Spatial Models}
\label{sec:source_and_spatoal_models}

MNMF~\cite{Sawada2013multichannel} and its constrained versions such as 
 ILRMA~\cite{kitamura2016determined} and FastMNMF~\cite{sekiguchi2020fast}
 are based on the low-rank source model
 that factorizes the PSDs of each source $n$ as
\begin{align}
\lambda_{nft} = \sum_{k=1}^{K}w_{nkf}h_{nkt},
\label{eq:nmf_source_gaussian}
\end{align} 
where $K$ is the number of bases, 
 $w_{nkf} \geq 0$ is the magnitude of basis $k$ of source $n$ at frequency $f$, 
 and $h_{nkt} \geq 0$ is the activation of basis $k$ of source $n$ at time $t$.
Let $\mbW \triangleq \{w_{nkf}\}_{n,k,f=1}^{N,K,F}$ 
 and $\mbH \triangleq \{h_{nkt}\}_{n,k,t=1}^{N,K,T}$
 be the sets of the bases and activations, respectively.
For ILRMA~\cite{kitamura2016determined},
 MNMF~\cite{Sawada2013multichannel},
 and FastMNMF~\cite{sekiguchi2020fast},
 the rank-1 spatial model,
 the unconstrained full-rank spatial model,
 and the jointly-diagonalizable full-rank spatial model have been proposed, respectively.

\subsubsection{Rank-1 Spatial Model}
\label{sec:rank_1}

Ideally, the sound propagation process in a less-echoic environment
 is represented as a time-invariant linear system as follows:
\begin{align}
 \mbx_{nft} &= \mba_{nf} s_{nft},
 \label{eq:x_nft}
\end{align}
where $\mba_{nf} \in \mathbb{C}^{M}$ is the steering vector
 of source $n$ at frequency $f$. 
Eq. \eqref{eq:x_nft} gives Eqs.~\eqref{eq:p_x_nft} and \eqref{eq:p_x_ft},
 where $\mbG_{nf} \triangleq \mba_{nf} \mba_{nf}^\Hr \succeq \bm{0} $ 
 is the rank-1 SCM of source $n$ at frequency $f$
 and ${}^\Hr$ denotes the conjugate transpose.

ILRMA~\cite{kitamura2016determined} is based on
 the low-rank source model given by Eq.~\eqref{eq:nmf_source_gaussian}
 and the rank-1 spatial model given by Eq.~\eqref{eq:p_x_ft} 
 with $\mbG_{nf} = \mba_{nf} \mba_{nf}^\Hr$.
It is available only under a determined condition ($M=N$)
 to avoid the rank deficiency of the SCM $\mbY_{ft}$ 
 for the observed mixture $\mbx_{ft}$.
 
\subsubsection{Full-Rank Spatial Model} 
\label{sec:full_rank}

Because Eq.~\eqref{eq:x_nft} does not hold
 when the reverberation is longer than the window size of STFT,
 one may want to allow $\mbG_{nf}$ to be a full-rank matrix~\cite{duong2010under}.
Note that Eqs.~\eqref{eq:p_x_nft} and \eqref{eq:p_x_ft} are not changed in form.

MNMF~\cite{Sawada2013multichannel} is based on the low-rank source model 
 given by Eq.~\eqref{eq:nmf_source_gaussian}
 and the full-rank spatial model given by Eq.~\eqref{eq:p_x_ft}
 with unconstrained $\mbG_{nf}$.
Unlike ILRMA,
 it can be used
 even under an underdetermined condition ($M < N$) in theory.
Because MNMF has a considerably larger number of spatial parameters than ILRMA ($NFM(M+1)/2 \gg NFM$),
 MNMF tends to easily get stuck in a bad local optimum.



\subsubsection{Jointly-Diagonalizable Spatial Model}

An effective way of reducing the complexity of MNMF 
 is to assume $\{\mbG_{nf}\}_{n=1}^N$ 
 to be jointly diagonalizable with a non-singular matrix $\mbQ_f \in \mathbb{C}^{M \times M}$
 called a \textit{diagonalizer}
 as follows \cite{Ito2018fastfca,Ito2018fastfcaas,ito2021joint,sekiguchi2020fast}:
\begin{align}
 \forall n,f, \  
 \mbG_{nf} = \mbQ_f^{-1} \mathrm{Diag}(\tilde{\mbg}_{nf}) \mbQ_f^{-\Hr}
 \quad \mbox{(version 1)},
 \label{eq:diagonalize1}
\end{align}
where $\tilde{\mbg}_{nf} \triangleq [\tilde{g}_{nf1}, \ldots, \tilde{g}_{nfM}]^{\Tr} \in \mathbb{R}_+^{M}$ 
 is a nonnegative vector of source $n$ at frequency $f$,
 $\mathrm{Diag}(\mbv)$ denotes a diagonal matrix 
 whose diagonal elements are given by a vector $\mbv$,
 and ${}^\Tr$ denotes the transpose.
Because $\mbQ_f \triangleq [\mbq_{f1}, \ldots, \mbq_{fM}]^\Hr \in \mathbb{C}^{M \times M}$ 
 acts as a demixing matrix consisting of $M$ demixing filters $\{\mbq_{fm}\}_{m=1}^M$,
 \ie, $\mbQ_f^{-1} \triangleq [\mbu_{f1},\cdots,\mbu_{fM}]$ 
 acts as a mixing matrix consisting of $M$ steering vectors $\{\mbu_{fm}\}_{m=1}^M$
 corresponding to different directions,
 $\tilde{\mbg}_{nf}$ is considered 
 to indicate the weights of the $M$ directions for source $n$. 
This naturally calls for sharing the direction weights 
 over all frequencies as follows:
\begin{align}
 \forall n,f, \  
 \mbG_{nf} = \mbQ_f^{-1} \mathrm{Diag}(\tilde{\mbg}_{n}) \mbQ_f^{-\Hr}
 \quad \mbox{(version 2)},
 \label{eq:diagonalize2}
\end{align}
where $\tilde{\mbg}_{n} \triangleq [\tilde{g}_{n1}, \ldots, \tilde{g}_{nM}]^{\Tr} \in \mathbb{R}_+^{M}$ 
 is a frequency-independent nonnegative vector of source $n$ \cite{sekiguchi2020fast}.
For better performance,
 we focus on this weight-shared version and define its diagonalizer set as $\mbQ \triangleq \{\mbQ_{f}\}_{f=1}^{F}$.
Note that the rank-1 spatial model is obtained 
 when $M=N$
 and $\tilde\mbG \triangleq [\tilde\mbg_1,\cdots,\tilde\mbg_N]^\Tr = \mbI$,
 where $\mbI$ denotes an identity matrix of size $M$.

FastMNMF2~\cite{sekiguchi2020fast} (simply called FastMNMF in this paper) 
 is obtained by integrating the low-rank source model 
 given by Eq.~\eqref{eq:nmf_source_gaussian}
 and the jointly-diagonalizable full-rank spatial model
 given by Eq.~\eqref{eq:p_x_ft} with Eq.~\eqref{eq:diagonalize2}.
Since the latent source image $\mbx_{nft}$ 
 and the observed mixture $\mbx_{ft}$ are Gaussian distributed,
 the \textit{projected} source $\mbz_{nft} \triangleq \mbQ_f \mbx_{nft}$ 
 and the \textit{projected} mixture $\mbz_{ft} \triangleq \mbQ_f \mbx_{ft}$ 
 are also Gaussian distributed as follows: 
\begin{align}
 \mbz_{nft}
 &\sim 
 \ComplexGaussian
 {\lambda_{nft} \mathrm{Diag}(\tilde{\mbg}_{n}) \triangleq \tilde\mbY_{nft}},
 \label{eq:p_z_nft}
 \\
 \mbz_{ft}
 &\sim 
 \ComplexGaussian
 {\sum_{n=1}^{N} \lambda_{nft} \mathrm{Diag}(\tilde{\mbg}_{n}) \triangleq \tilde\mbY_{ft}},
 \label{eq:p_z_ft}
\end{align}
MNMF for $\mbz_{ft}$ is thus a particular case of nonnegative tensor factorization (NTF) 
 that assumes the elements of $\mbz_{ft}$ to be independent,
 whereas those of $\mbx_{ft}$ are correlated. (see Fig.~2 in \cite{sekiguchi2020fast}).



\subsection{Gaussian and Heavy-Tailed Models}\label{sec:ex_FastMNMF}
 
We explain the probabilistic model of FastMNMF 
 (called $\mathcal{N}$-FastMNMF~\cite{sekiguchi2020fast})
 and those of the Student's $t$
 and leptokurtic GG extensions of $\mathcal{N}$-FastMNMF
 that can handle more impulsive sources.
Such an extension is achieved by replacing the Gaussian distribution
 with a surrogate distribution in Eq.~\eqref{eq:p_z_ft}.
Let $\mbTh \triangleq \{\mbW, \mbH, \mbQ, \tilde\mbG\}$
 be a set of model parameters.

\subsubsection{Gaussian FastMNMF}
\label{sec:Gaussian-FastMNMF}

Using the change-of-variable principle for $\mbz_{ft} = \mbQ_f \mbx_{ft}$,
 the log-likelihood (LL) of the parameters $\mbTh$
 for the observed mixture $\mbX$ is given by 
\begin{align}
  \!\!
  \log p_{\mbTh}(\mbX)
  &=
  \sum_{f,t=1}^{F,T} \log p(\mbz_{ft}) 
  + \sum_{f,t=1}^{F,T} \log \left|\frac{d \mbz_{ft}}{d \mbx_{ft}}\right|
  \nonumber\\
  &=
  \sum_{f,t=1}^{F,T} \log p(\mbz_{ft}) 
  + T \sum_{f=1}^F \log \left| \mbQ_{f} \mbQ_{f}^{\Hr} \right|,
  \label{eq:log_likelihood_jdsm}
\end{align}
where $\log p(\mbz_{ft})$ is given by
\begin{align}
  \log p(\mbz_{ft})
  \protect\overset{c}{=}
  - \sum_{m=1}^{M} \frac{\tilde{z}_{ftm}}{\tilde{y}_{ftm}} 
  - \sum_{m=1}^{M} \log \tilde{y}_{ftm},
  \label{eq:log_likelihood_jdsm_gaussian}
\end{align}
 where $\overset{c}{=}$ denotes
 equality up to an additive constant 
 and 
\begin{align}
 \tilde{z}_{ftm}
 &\triangleq 
 \left|z_{ftm}\right|^2
 =
 \left|\mbq_{fm}^\Hr \mbx_{ft}\right|^2,
 \label{eq:xti_ftm} \\
 \tilde{y}_{ftm} 
 &\triangleq 
 \sum_{n=1}^{N} \lambda_{nft} \tilde{g}_{nm} 
 =
 \sum_{n,k=1}^{N, K} w_{nkf} h_{nkt} \tilde{g}_{nm}.
 \label{eq:yti_ftm}
\end{align}

\subsubsection{Student's $t$ FastMNMF}
\label{sec:t-FastMNMF}

$t$-FastMNMF~\cite{kamo2020jointstudent}
 with a degree of freedom $\nu>0$ controlling the tail lightness
 reduces to $\mathcal{N}$-FastMNMF \cite{sekiguchi2020fast} when $\nu \rightarrow \infty$,
 and reduces to $t$-R1-FastMNMF when the rank-1 spatial model is used.
More specifically, Eq.~\eqref{eq:p_z_ft} is replaced with
\begin{align}
 \mbz_{ft} \sim \ComplexStudent{\nu}
 {\tilde\mbY_{ft}},
 \label{eq:p_z_ft_student}
\end{align}
where $\mathcal{T}^{\nu}_{\mathbb{C}}(\mbSi)$ 
 denotes a zero-mean multivariate complex $t$ distribution 
 with a degree of freedom $\nu>0$ and a scale matrix $\mbSi \succeq \bm{0}$
 (the PDF is given by Eq.~\eqref{eq:pdf-T}).
The $t$ distribution approaches the Gaussian distribution as $\nu \rightarrow \infty$.
For reference,
the real parts of univariate complex $t$ distributions are plotted in Fig.~\ref{fig:GSM-pdf}.
The LL of the parameters $\mbTh$ is the same in form as Eq.~\eqref{eq:log_likelihood_jdsm},
 where $\log p(\mbz_{ft})$ is given by
\begin{align}
    &\!\!\!\!
    \log p(\mbz_{ft})
    \nonumber\\
    &\!\!\!\!
    \protect\overset{c}{=}
    -
    \left(\frac{\nu}{2} + M\right)
    \log\!
    \left(1+\frac{2}{\nu}\sum_{m=1}^M\frac{\tilde{z}_{ftm}}{\tilde{y}_{ftm}}\right)
    - \sum_{m=1}^M \log \tilde{y}_{ftm}.
    \!
    \label{eq:log_likelihood_jdsm_t}
\end{align}

\subsubsection{Leptokurtic Generalized Gaussian FastMNMF}
\label{sec:GG-FastMNMF}

Leptokurtic GG-FastMNMF
 with a shape parameter $\beta \in (0, 2]$ controlling the tail lightness
 reduces to $\mathcal{N}$-FastMNMF \cite{sekiguchi2020fast} when $\beta = 2$,
 and reduces to leptokurtic GG-R1-FastMNMF with $\beta \in (0, 2]$
 when the rank-1 spatial model is used.
Note that leptokurtic GG-FastMNMF with $\beta \in (0, 2]$
 has not been investigated in the literature,
 whereas platykurtic GG-FastMNMF~\cite{kamo2020jointsubgaussian} 
 and its ILRMA version ~\cite{mogami2019independent} with $\beta \in [2, 4)$
 have already been proposed.
More specifically, Eq.~\eqref{eq:p_z_ft} is replaced with
\begin{align}
 \mbz_{ft} \sim \ComplexGG{\beta}{\tilde\mbY_{ft}},
 \label{eq:p_z_ft_GG}
\end{align}
where $\ComplexGG{\beta}{\mbSi}$ 
 denotes a zero-mean leptokurtic multivariate complex GG distribution \cite{palmer2006variational} 
 with a shape parameter $\beta \in (0, 2]$ 
 and a scale matrix $\mbSi \succeq \bm{0}$
 (the PDF is given by Eq.~\eqref{eq:pdf-GG}).
The GG distribution with $\beta=2$
 reduces to the Gaussian distribution.
For reference,
 the real parts of leptokurtic univariate complex GG distributions are plotted in Fig.~\ref{fig:GSM-pdf}.
The LL of the parameters $\mbTh$
 is the same in form as Eq.~\eqref{eq:log_likelihood_jdsm},
 where $\log p(\mbz_{ft})$ is given by
\begin{align}
 & \log p(\mbz_{ft})
 \protect\overset{c}{=}
 -
 \left(\sum_{m=1}^M\frac{\tilde{z}_{ftm}}{\tilde{y}_{ftm}}\right)^{\frac{\beta}{2}}
 -
 \sum_{m=1}^M\log \tilde{y}_{ftm}.
 \label{eq:log_likelihood_jdsm_gg}
\end{align}



\subsubsection{$\alpha$-Stable FastMNMF}
\label{sec:alpha-FastMNMF}

\begin{figure}
\centering
\includegraphics[width=.99\columnwidth]{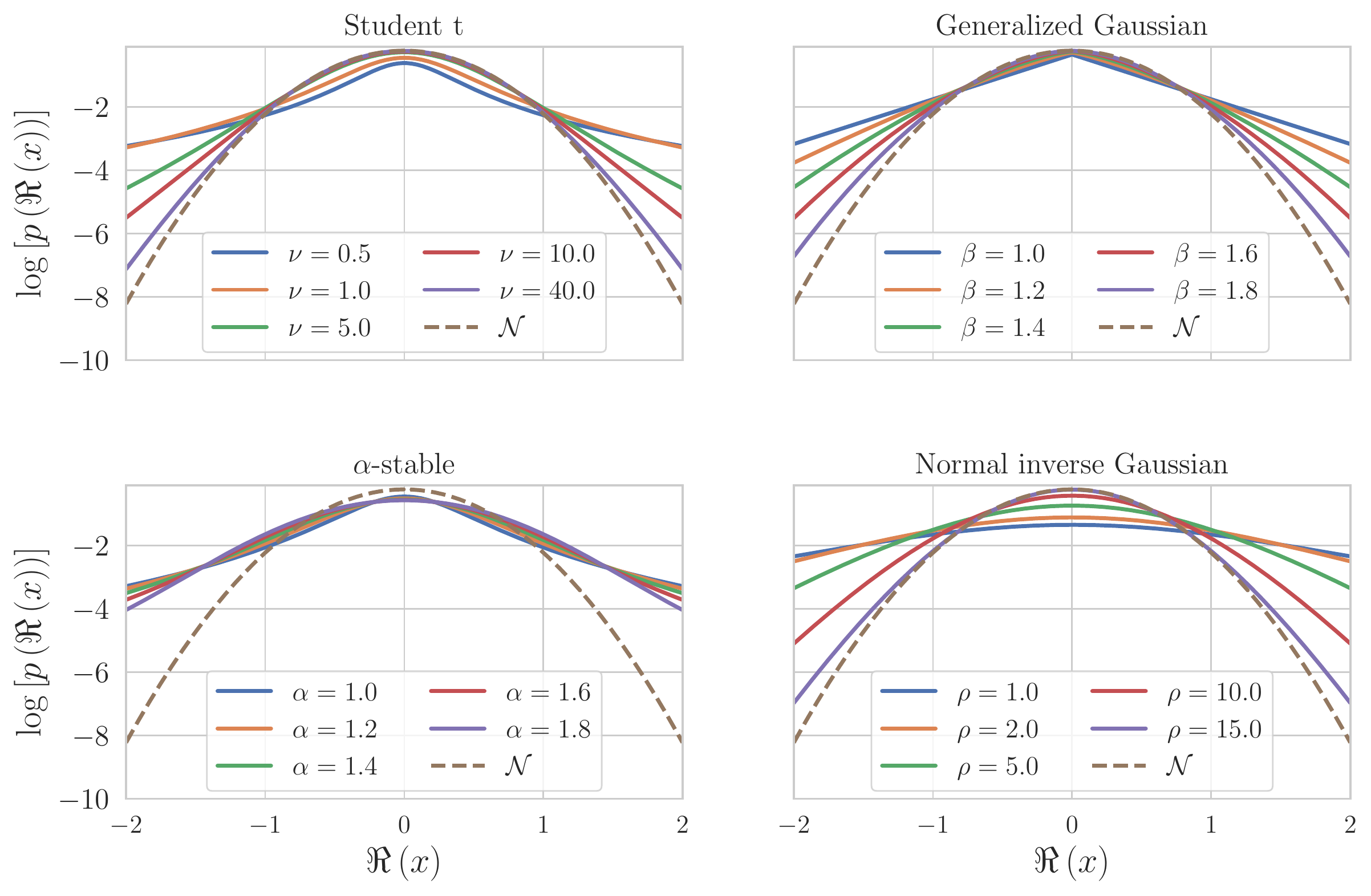}
\vspace{-2mm}
\caption{
Standard univariate complex GSMs on the real axis.
Top left: Student's $t$ distributions
 with degrees of freedom $\nu > 0$.
Top right: Leptokurtic generalized Gaussian (GG) distributions with shape parameters $\beta \in (0, 2]$.
Bottom left: $\alpha$-stable distributions with characteristic exponents $\alpha \in (0, 2]$.
Bottom right: Normal inverse Gaussian (NIG) distributions with concentration parameters $\rho >0$.
}
\label{fig:GSM-pdf}
\end{figure}

$\alpha$-FastMNMF~\cite{fontaine2020unsupervised}
 with a characteristic exponent $\alpha \in [0, 2)$ controlling the tail lightness
 reduces to $\mathcal{N}$-FastMNMF \cite{sekiguchi2020fast} when $\alpha = 2$,
 and reduces to $\alpha$-R1-FastMNMF~\cite{fontaine2021alpha} 
 when the rank-1 spatial model is used.
More specifically, Eq.~\eqref{eq:p_z_ft} is replaced with
\begin{align}
 \mbz_{ft} 
 \sim 
 \ComplexStable{\alpha}{\tilde\mbY_{ft}},
 \label{eq:p_z_ft_stable}
\end{align}
where $\mathcal{S}^{\alpha}_{\mathbb{C}}(\mbSi)$ 
 denotes a zero-mean non-skewed multivariate elliptically complex $\alpha$-stable distribution 
 with a characteristic exponent $\alpha>0$ and a scale matrix $\mbSi \succeq \bm{0}$ \cite{leglaive2017alpha}.
For reference,
 the real parts of univariate complex $\alpha$-stable distributions are plotted in Fig.~\ref{fig:GSM-pdf}.
The LL of the parameters $\mbTh$
 is the same in form as Eq.~\eqref{eq:log_likelihood_jdsm},
 where in general $\log p(\mbz_{ft})$ cannot be expressed
 in a closed form
 except for $\alpha \in \{\frac{1}{2}, 1, 2\}$,
 making ML estimation of $\mbTh$ challenging.
To circumvent this problem,
 one can rewrite Eq.~\eqref{eq:p_z_ft_stable}
 as an analytically-tractable GSM representation (cf. Section~\ref{sec:GSM-FastMNMF}),
 where the auxiliary impulse variable needs to be marginalized out
 with a computationally-expensive MH algorithm
 \cite{fontaine2020unsupervised}.
Note that $\alpha$-FastMNMF is not dealt with in this paper
 because deterministic parameter update rules cannot be obtained.

\section{Gaussian Scale Mixture Fast Multichannel Nonnegative Matrix Factorization}
\label{sec:GSM-FastMNMF}

We propose GSM-FastMNMF, a general form of heavy-tailed FastMNMF,
 including $\mathcal{N}$-FastMNMF~\cite{sekiguchi2020fast} (Section~\ref{sec:Gaussian-FastMNMF}),
 its heavy-tailed extensions such as
 $t$-FastMNMF~\cite{kamo2020jointstudent} (Section~\ref{sec:t-FastMNMF}),
 leptokurtic GG-FastMNMF (Section~\ref{sec:GG-FastMNMF}),
 and $\alpha$-FastMNMF~\cite{fontaine2020unsupervised} (Section~\ref{sec:alpha-FastMNMF})
 and the rank-1 counterparts such as 
 $t$-R1-FastMNMF,
 leptokurtic GG-R1-FastMNMF,
 and $\alpha$-R1-FastMNMF~\cite{fontaine2021alpha}.
The closed-form deterministic parameter update rules
 have been tailor-made independently for the existing variants
 except for $\alpha$-FastMNMF based on the stochastic parameter update rules with the MH sampler.

We explain a probabilistic model of GSM-FastMNMF
 and derive its parameter estimation algorithm.
As a concrete example of GSM-FastMNMF,
 we then instantiate GH-FastMNMF based on the generalized hyperbolic (GH) distribution
 as a wide family of heavy-tailed FastMNMF 
 including $\mathcal{N}$-FastMNMF~\cite{sekiguchi2020fast} and
 $t$-FastMNMF~\cite{kamo2020jointstudent}.
As a well-performing special case of GH-FastMNMF,
 we focus on NIG-FastMNMF based on the normalized inverse Gaussian (NIG) distribution.

\subsection{Probabilistic Formulation}
\label{sec:GSM}



%
GSM-FastMNMF is obtained 
 by extending the multivariate complex Gaussian distributions 
 used in Eqs.~\eqref{eq:p_z_nft} and \eqref{eq:p_z_ft} 
 to multivariate complex GSMs represented 
 as compound probability distributions as follows:
\begin{align}
\phi_{ft}
&\sim
p(\phi_{ft}),
\label{eq:p_phi_ft_gsm}
\\
\mathbf{z}_{nft}\mid\mbTh, \phi_{ft}
&\sim
\ComplexGaussian{\phi_{ft} \tilde\mbY_{nft}},
\label{eq:p_z_nft_gsm}
\\
\mathbf{z}_{ft}\mid\mbTh,\phi_{ft}
&\sim
\ComplexGaussian{\phi_{ft} \tilde\mbY_{ft}},
\label{eq:p_z_ft_gsm}
\end{align}
with $\phi_{ft} > 0$ is an auxiliary nonnegative random variable called an impulse variable 
 that stochastically perturbs the covariance matrices $\tilde\mbY_{nft}$ and $\tilde\mbY_{ft}$
 according to some prior distribution $p(\phi_{ft})$.
The LL of the parameters $\mbTh$ is given by
\begin{align}
 \log p_{\mbTh}(\mbX) = \log \int p_{\mbTh}(\mbX \mid \mbPh) p(\mbPh) d\mbPh,
 \label{eq:log_likelihood_jdsm_gsm}
\end{align}
where $\mbPh \triangleq \{\phi_{ft}\}_{f,t=1}^{F,T}$ and
 $p_{\mbTh}(\mbX \mid \mbPh)$ is the same in form as Eq.~\eqref{eq:log_likelihood_jdsm}
 except that the Gaussian density $p(\mbz_{ft})$ is replaced with
 the \textit{conditional} Gaussian density $p(\mbz_{ft} \mid \phi_{ft})$
 given by
\begin{align}
  \log p(\mbz_{ft} \mid \phi_{ft})
  \protect\overset{c}{=}
  - \sum_{m=1}^{M} \frac{\tilde{z}_{ftm}}{\phi_{ft} \tilde{y}_{ftm}} 
  - \sum_{m=1}^{M} \log \phi_{ft} \tilde{y}_{ftm}.
  \label{eq:log_likelihood_jdsm_conditional_gaussian}
\end{align}
Note that several existing heavy-tailed extensions of FastMNMF
 are obtained by marginalizing $\mbPh$ out with the mixing distribution $p(\mbPh)$
 according to Eq.~\eqref{eq:log_likelihood_jdsm_gsm}.

\subsection{Multiplicative Update Variational Expectation-Maximization Algorithm}

 We describe in that Section how parameters $\mbTh$ are estimated.
Since the LL of $\mbTh$, $\log p_{\mbTh}(\mbX)$, 
 given by Eq.~\eqref{eq:log_likelihood_jdsm_gsm}
 is hard to directly maximize with respect to $\mbTh$, 
 we use a multiplicative update variational expectation-maximization (MU-VEM) principle,
 \ie, derive a variational lower bound 
 $\mathcal{L}(\mbTh,q(\mbPh),\mbPs)$ of $\log p_{\mbTh}(\mbX)$ 
 using an arbitrary distribution $q(\mbPh)$ 
 of the latent impulse variables $\mbPh$
 and a set of auxiliary variables $\mbPs$ (Section \ref{sec:elbo})
 and iteratively
 update $q(\mbPh)$ and $\mbPs$ in the E-step (Section \ref{sec:E-Step}) 
 and $\mbTh$ in the M-step (Section \ref{sec:M-Step})
 such that $\mathcal{L}(\mbTh,q(\mbPh),\mbPs)$ monotonically non-decreases.

\subsubsection{Lower Bound}
\label{sec:elbo}

Let $q(\mbPh) \triangleq \prod_{f,t=1}^{F,T} q(\phi_{ft})$ be 
 an arbitrary distribution on the latent impulse variables $\mbPh$.
Using Jensen's inequality,
 Eq.~\eqref{eq:log_likelihood_jdsm_gsm} can be lower bounded as follows:
\begin{align}
\log p_{\mbTh}(\mbX) 
 &=
 \sum_{f,t=1}^{F,T}
 \log\int p_{\mbTh}(\mbx_{ft}\mid\phi_{ft})p(\phi_{ft})d\phi_{ft}
 \nonumber\\
 &=
 \sum_{f,t}^{}
 \log\int q(\phi_{ft}) \frac{p_{\mbTh}(\mbx_{ft}\mid\phi_{ft})p(\phi_{ft})}{q(\phi_{ft})}d\phi_{ft}
 \nonumber\\
 & \geq
 \sum_{f,t}^{}
 \Big(
 \mathbb{E}_{q(\phi_{ft})}\left[\log p_{\mbTh}(\mbz_{ft}\mid\phi_{ft})\right]
 + \left|\mbQ_{f}\mbQ_{f}^{\Hr}\right|
 \nonumber\\
 &\quad\quad\quad\quad
 -\mathrm{KL}\left[q(\phi_{ft}) \parallel p(\phi_{ft})\right]
 \Big)
 \nonumber\\
 &\triangleq
 \mathcal{L}'(\mbTh, q(\mbPh)),
\end{align}
where 
 $\mathrm{KL}(q \parallel p)$ denotes the Kullback-Leibler (KL) divergence
 from $q$ to $p$~\cite{kullback1951information},
 and $p_{\mbTh}(\mbz_{ft}\mid\phi_{ft})$ is given 
 by Eq.~\eqref{eq:log_likelihood_jdsm_conditional_gaussian}.
The equality condition that maximizes $\mathcal{L}'(\mbTh,q(\mbPh))$ is given by
\begin{align}
q(\phi_{ft}) 
= 
p(\phi_{ft} \mid \mbx_{ft})
= 
p(\phi_{ft} \mid \mbz_{ft}).
\label{eq:eq_cond_q}
\end{align}

Let $\mbPs \triangleq \{\bm\Pi,\bm\Omega\}$
 be a set of arbitrary nonnegative variables,
 where $\bm\Pi \triangleq \{\pi_{ftmnk}\}_{f,t,m,n,k=1}^{F,T,M,N,K}$
 satisfying $\sum_{n,k=1}^{N,K} \pi_{ftmnk} = 1$
 and $\bm\Omega \triangleq \{\omega_{ftm}\}_{f,t,m=1}^{F,T,M}$.
Since $\mathcal{L}'(\mbTh, q(\mbPh))$ is still hard to maximize with respect to $\mbTh$, 
 it is further lower bounded as in NMF based on the Itakura-Saito (IS) divergence\cite{nakano2010convergence} as follows:
\begin{align}
 \mathcal{L}'(\mbTh, q(\mbPh))
 &=
 - \!\!\sum_{f,t,m=1}^{F,T,M}\!\!
 \Bigg(
 \frac{\mathbb{E}_{q(\phi_{ft})}\big[\phi_{ft}^{-1}\big]\tilde{z}_{ftm}}
 {\sum_{n,k=1}^{N, K} w_{nkf} h_{nkt} \tilde{g}_{nm}} 
 \nonumber\\
 &\quad\quad\quad\quad\quad\quad 
 + \mathbb{E}_{q(\phi_{ft})}[\log \phi_{ft}] 
 \nonumber\\
 &\quad\quad\quad\quad\quad\quad
 + \log\!\Bigg(\sum_{n,k=1}^{N, K} w_{nkf} h_{nkt} \tilde{g}_{nm}\Bigg) 
 \Bigg)
 \nonumber\\
 &\quad
 + \sum_{f,t=1}^{F,T}
 \Big(
 \left|\mbQ_{f}\mbQ_{f}^{\Hr}\right|
 -\mathrm{KL}\left[q(\phi_{ft}) \parallel p(\phi_{ft})\right]
 \Big)
 \nonumber\\
 &\geq
 - \sum_{f,t,m}
 \Bigg(
 \sum_{n,k}
 \frac{\mathbb{E}_{q(\phi_{ft})}\big[\phi_{ft}^{-1}\big] 
 \pi_{nftmk}^{2}  \tilde{z}_{ftm}}
 {w_{nkf} h_{nkt} \tilde{g}_{nm}}
 \nonumber\\
 &\quad\quad\quad\quad
 + \mathbb{E}_{q(\phi_{ft})}[\log \phi_{ft}] 
 \nonumber\\
 &\quad\quad\quad\quad
 + \log \omega_{ftm} + \sum_{n,k} \frac{w_{nkf} h_{nkt} \tilde{g}_{nm}}{\omega_{ftm}} - 1
 \Bigg)
 \nonumber\\
 &\quad
 + \sum_{f,t}^{}
 \Big(
 \left|\mbQ_{f}\mbQ_{f}^{\Hr}\right|
 -\mathrm{KL}\left[q(\phi_{ft}) \parallel p(\phi_{ft})\right]
 \Big)
 \nonumber\\
 &\triangleq
 \mathcal{L}(\mbTh,q(\mbPh),\mbPs).
 \label{eq:LL-lower-bound}
\end{align}
Letting the partial derivative of Eq.~\eqref{eq:LL-lower-bound}
 with respect to $\mbPs$ equal to zero,
 the equality condition that maximizes $\mathcal{L}(\mbTh,q(\mbPh),\mbPs)$
 is given by
\begin{align}
\pi_{ftmnk} &= w_{nkf} h_{nkt} \tilde{g}_{nm} \tilde{y}_{ftm}^{-1},
\label{eq:eq_cond_pi}
\\
\omega_{ftm} &= \tilde{y}_{ftm}.
\label{eq:eq_cond_omega}
\end{align}


\subsubsection{E-Step}
\label{sec:E-Step}

Given the current estimate of $\mbTh$,
 we update $q(\mbPh)$ using Eq.~\eqref{eq:eq_cond_q}
 and $\mbPs$ using Eqs.~\eqref{eq:eq_cond_pi} and \eqref{eq:eq_cond_omega}
 such that the lower bound $\mathcal{L}(\mbTh,q(\mbPh),\mbPs)$ 
 given by Eq.~\eqref{eq:LL-lower-bound} is maximized
 with respect to $q(\mbPh)$ and $\mbPs$.
Note that the optimal estimate of $q(\phi_{ft})$ given by Eq.~\eqref{eq:eq_cond_q}
 is used for computing the posterior expectation 
 $\mathbb{E}_{q(\phi_{ft})}\big[\phi_{ft}^{-1}\big]$
 used in the M-step.
The tractability of
 $\tilde{\phi}_{ft}^{-1} \triangleq 
 \mathbb{E}_{p(\phi_{ft} \mid \mbz_{ft})}\big[\phi_{ft}^{-1}\big]$
 is thus a key for deriving closed-form update rules.
Let $\tilde\mbPh \triangleq \{\tilde{\phi}_{ft}\}_{f,t=1}^{F,T}$
 be a set of the posterior expectations.
As derived in the Appendix,
 we have
\begin{align}
\frac{d}{d\mbz_{ft}^{\Hr}} \log p(\mbz_{ft}) 
=
- 2\tilde{\phi}_{ft}^{-1}\tilde{\mbY}_{ft}^{-1} \mbz_{ft}
\label{eq:e_log_p_zft}
\end{align} where $\tilde{\mbY}_{ft}$ is defined in Eq.~\eqref{eq:p_z_ft}.
Note that 
 even if the posterior density $p(\phi_{ft} \mid \mbz_{ft})$ is intractable,
 $\tilde{\phi}_{ft}^{-1}$ is tractable 
 if the log-marginal density $\log p(\mbz_{ft})$ is differentiable 
 with respect to $\mbz_{ft}$ (\eg, GG-FastMNMF).



\subsubsection{M-Step}\label{sec:M-Step}

Given the current estimates of $q(\mbPh)$ and $\mbPs$,
 we update $\mbTh$ such that the lower bound $\mathcal{L}(\mbTh,q(\mbPh),\mbPs)$ 
 given by Eq.~\eqref{eq:LL-lower-bound} is maximized
 with respect to $\mbTh$,
 in the same way as $\mathcal{N}$-FastMNMF~\cite{sekiguchi2020fast}.
Letting the partial derivative of Eq.~\eqref{eq:LL-lower-bound}
 with respect to $\mbW$, $\mbH$, and $\tilde\mbG$ equal to zero
 and using Eq.~\eqref{eq:eq_cond_q}, 
 \eqref{eq:eq_cond_pi}, and \eqref{eq:eq_cond_omega},
 the update rules of $\mbW$, $\mbH$, and $\tilde\mbG$
 are obtained in a closed form as follows:
\begin{align}
w_{nkf} & \leftarrow w_{nkf}\sqrt{\frac{\sum_{t,m=1}^{T,M}h_{nkt}\tilde{g}_{nm}\tilde{y}_{ftm}^{-2}\hat{z}_{ftm}}{\sum_{t,m=1}^{T,M}h_{nkt}\tilde{g}_{nm}\tilde{y}_{ftm}^{-1}}},\label{eq:update_W} \\
h_{nkt} & \leftarrow h_{nkt}\sqrt{\frac{\sum_{f,m=1}^{F,M}w_{nkf}\tilde{g}_{nm}\tilde{y}_{ftm}^{-2}\hat{z}_{ftm}}{\sum_{f,m=1}^{F,M}w_{nkf}\tilde{g}_{nm}\tilde{y}_{ftm}^{-1}}},\label{eq:update_H} \\
\tilde{g}_{nm} & \leftarrow\tilde{g}_{nm}\sqrt{\frac{\sum_{f,t=1}^{F,T}\lambda_{nft}\tilde{y}_{ftm}^{-2}\hat{z}_{ftm}}{\sum_{f,t=1}^{F,T}\lambda_{nft}\tilde{y}_{ftm}^{-1}}},\label{eq:update_G}
\end{align}
where $\hat{z}_{ftm}$ is given by
\begin{align}
\hat{z}_{ftm} 
= 
\tilde{\phi}_{ft}^{-1} \tilde{z}_{ftm}. 
\label{eq:zhat_ftm}
\end{align}
The update rule of $\mbQ$ is also obtained in a closed form
 with iterative projection (IP)~\cite{ono2011stable} as follows:
\begin{align}
\mbV_{fm} & \triangleq\frac{1}{T}\sum_{t=1}^{T} \tilde{\phi}_{ft}^{-1} \mbX_{ft} \tilde{y}_{ftm}^{-1},
\\
\mbq_{fm} & \leftarrow\left(\mbQ_{f}\mbV_{fm}\right)^{-1}\mbe_{m},
\label{eq:update_Q1}\\
\mbq_{fm} & \leftarrow\left(\mbq_{fm}^{\mathrm{\mathsf{H}}}\mbV_{fm}\mbq_{fm}\right)^{-\frac{1}{2}} \mbq_{fm},
\label{eq:update_Q2}
\end{align}
where $\mbe_m$ is a one-hot vector whose $m$-th entry is $1$ and $0$ elsewhere.
To avoid scale ambiguity, the parameters are normalized as follows:
\begin{align}
 r_f &= M\mathrm{Tr}\left(\mbQ_{f}\mbQ_{f}^{\Hr}\right),
 &
 \begin{cases}
 \mbQ_{f} \leftarrow r_f^{-\frac{1}{2}} \mbQ_{f},
 \\
 w_{nkf} \leftarrow r_f^{-1} w_{nkf},
 \end{cases}
 \label{eq:norm1} 
 \\
 u_n &= \sum_{m=1}^{M}\tilde{g}_{nm},
 &
 \begin{cases}
 \tilde{g}_{nm} \leftarrow u_n^{-1} \tilde{g}_{nm},
 \\
 w_{nkf} \leftarrow u_n w_{nkf}.
 \end{cases}
 \label{eq:norm2}
 \\
 v_{nk} &= \sum_{f=1}^{F}w_{nkf},
 &
 \begin{cases}
 w_{nkf} \leftarrow v_{nk}^{-1} w_{nkf},
 \\
 h_{nkt} \leftarrow v_{nk} h_{nkt}.
 \end{cases}
 \label{eq:norm3}
\end{align}


\subsection{Existing Instances of GSM-FastMNMF}

We show that $\mathcal{N}$-FastMNMF (Section~\ref{sec:Gaussian-FastMNMF}),
 $t$-FastMNMF (Section~\ref{sec:t-FastMNMF}),
 leptokurtic GG-FastMNMF (Section~\ref{sec:GG-FastMNMF}),
 and $\alpha$-FastMNMF (Section~\ref{sec:alpha-FastMNMF})
 can readily be instantiated from GSM-FastMNMF.
The update rules of the parameters $\mbTh = \{\mbW, \mbH, \mbQ, \tilde\mbG\}$
 are commonly given by Eqs.~\eqref{eq:update_W}--\eqref{eq:norm3}
 and the posterior expectations $\tilde\mbPh$
 can be computed using Eq.~\eqref{eq:e_log_p_zft}.
For each model, 
 we instantiate the mixing distribution $p(\phi_{ft})$ given by Eq.~\eqref{eq:p_phi_ft_gsm}
 and compute $\tilde\phi_{ft}$ and $\hat{z}_{ftm}$ 
 according to Eqs.~\eqref{eq:e_log_p_zft} and \eqref{eq:zhat_ftm}, respectively.

\subsubsection{Gaussian FastMNMF}
\label{sec:gaussian_instance}

$\mathcal{N}$-FastMNMF~\cite{sekiguchi2020fast} 
 is obtained when $\phi_{ft} = 1$, \ie,
\begin{align}
 \phi_{ft} \sim \delta(\phi_{ft} - 1),
 \label{eq:phi_Gauss}
\end{align}
where $\delta(x)$ is the Dirac's delta function taking infinity at $x = 0$ and zero otherwise.
In this case, 
 Eq.~\eqref{eq:p_z_ft_gsm} reduces to Eq.~\eqref{eq:p_z_ft}.
Using Eq.~\eqref{eq:log_likelihood_jdsm_gaussian}
 and Eq.~\eqref{eq:e_log_p_zft}, we have
\begin{align}
 \tilde\phi_{ft}^{-1} = 1.
 \label{eq:e_inv_phi_ft_gaussian}
\end{align} 

\subsubsection{Student's $t$ FastMNMF}
\label{sec:t_instance}

$t$-FastMNMF~\cite{kamo2020jointstudent} 
 with a degree of freedom $\nu>0$ is obtained when $\phi_{ft}$ follows an inverse gamma (IG) distribution, denoted $\mathcal{IG}\!\left(a, b\right)$ where $a>0$ is a shape parameter and $b>0$ is a scale parameter, and by setting $a=b=\frac{\nu}{2}$ (see Eq.~\eqref{eq:pdf_IG} for the PDF): 
\begin{align}
\phi_{ft} \sim \mathcal{IG}\!\left(\frac{\nu}{2}, \frac{\nu}{2}\right), \label{eq:phi_ft_IG}
\end{align} 

The marginalization of $\phi_{ft}$ with Eqs.~\eqref{eq:p_phi_ft_gsm} and \eqref{eq:p_z_ft_gsm}
 gives Eq.~\eqref{eq:p_z_ft_student}.
Using Eq.~\eqref{eq:log_likelihood_jdsm_t}
 and Eq.~\eqref{eq:e_log_p_zft}, we have
\begin{align}
 \tilde\phi_{ft}^{-1}
 =
 \frac
 {\frac{\nu}{2} + M}
 {\frac{\nu}{2} + \sum_{m=1}^{M} \frac{\tilde{z}_{ftm}}{\tilde{y}_{ftm}}}.
 \label{eq:e_inv_phi_ft_student}
\end{align} 
$t$-FastMNMF with Eq.~\eqref{eq:e_inv_phi_ft_student} 
 approaches $\mathcal{N}$-FastMNMF with Eq.~\eqref{eq:e_inv_phi_ft_gaussian}
 as $\nu$ diverges to infinity.


\subsubsection{Leptokurtic Generalized Gaussian FastMNMF}
\label{sec:gg_instance}

Leptokurtic GG-FastMNMF
 with a shape parameter $\beta \in (0, 2]$ is known to be a GSM,
 but $p(\phi_{ft})$ is related to a positive $\alpha$-stable distribution
 whose PDF cannot be represented in a closed form
 except for the Gaussian case ($\beta=2$).
Nonetheless, using Eq.~\eqref{eq:log_likelihood_jdsm_gg}
 and Eq.~\eqref{eq:e_log_p_zft}, we have
\begin{align}
 \tilde{\phi}_{ft}^{-1}
 =
 \frac{\beta}{2}\!
 \left(\sum_{m=1}^{M}\frac{\tilde{z}_{ftm}}{\tilde{y}_{ftm}}\right)^{\frac{\beta-2}{2}}.
 \label{eq:e_inv_phi_ft_gg}
\end{align}
GG-FastMNMF with Eq.~\eqref{eq:e_inv_phi_ft_gg} 
 reduces to $\mathcal{N}$-FastMNMF with Eq.~\eqref{eq:e_inv_phi_ft_gaussian}
 when $\beta=2$.
When $\beta=2$ for GG-FastMNMF in Eq.~\eqref{eq:e_inv_phi_ft_gg}, it implies that $\forall (f,t), \tilde{\phi}_{ft}^{-1} = 1$ which describes the MUs of $\mathcal{N}$-FastMNMF. 
\subsubsection{$\alpha$-Stable FastMNMF}

$\alpha$-FastMNMF~\cite{fontaine2020unsupervised}
 with a characteristic exponent $\alpha \in [0, 2)$ 
 is obtained when $\phi_{ft}$ follows a positive $\frac{\alpha}{2}$-stable distribution, denoted $\mathcal{S}^\alpha_{\mathbb{R}+}(v)$ where $v > 0$ is a scale parameter, and by setting $v=2\cos\left(\frac{\pi\alpha}{4}\right)^{\frac{2}{\alpha}}$:
\begin{align}
\phi_{ft} \sim \mathcal{S}^\alpha_{\mathbb{R}+}\!
\left(2\cos\left(\frac{\pi\alpha}{4}\right)^{\frac{2}{\alpha}}\right),\label{eq:phi_ft_Palpha}
\end{align} 
The marginalization of $\phi_{ft}$ with Eqs.~\eqref{eq:p_phi_ft_gsm} and \eqref{eq:p_z_ft_gsm}
 gives Eq.~\eqref{eq:p_z_ft_stable}.
In general, the PDF of the $\alpha$-stable distribution 
 has no closed-form expression except 
 for the Levy ($\alpha=\frac{1}{2}$), Cauchy ($\alpha=1$), and Gaussian ($\alpha=2$) cases.
It is thus necessary to approximately compute $\tilde{\phi}_{ft}^{-1}$
 using an MH sampler as in \cite{fontaine2020unsupervised}.
Investigation of the existence and derivation of a closed-form expression of $\tilde{\phi}_{ft}^{-1}$
 remains as future work.
 
\subsection{Generalized Hyperbolic FastMNMF}
\label{sec:GH-FastMNMF}

We propose a new instance of GSM-FastMNMF 
 based on the multivariate complex generalized hyperbolic (GH) likelihood
 (denoted by $\mathcal{GH}_{\mathbb{C}}^{\gamma, \rho, \eta}$), called GH-FastMNMF.
Its constrained version called GH-R1-FastMNMF is obtained
 when the rank-1 spatial model is used.
The multivariate GH distribution~\cite{barndorff1977infinite}
 has infinite divisibility property~\cite{podgorski2016convolution},
 \ie, a GH random vector can be decomposed 
 into the sum of i.i.d. random vectors~\cite{ken1999levy}.
Since the GH distribution is closed under affine transformation,
 it has high affinity 
 to the joint diagonalizability of FastMNMF given by Eq.~\eqref{eq:diagonalize2}
 because the observed mixture $\mbx_{ft}$ 
 following a GH distribution with a \textit{full} scale matrix
 can be transformed to the projected mixture $\mbz_{ft} = \mbQ_f \mbx_{ft}$
 following a GH distribution with a \textit{diagonal} scale matrix.
 
\subsubsection{Probabilistic Formulation}

GH-FastMNMF
 is obtained by replacing Eq.~\eqref{eq:p_z_ft} with (see Eq.~\eqref{eq:pdf-GH} for the PDF)
\begin{align}
 \mbox{GH-FastMNNF:}\
 \mbz_{ft}
 \sim
 \ComplexGH{\gamma,\rho,\eta}{\tilde\mbY_{ft}},
 \label{eq:p_z_ft_gh}
\end{align}
where $\ComplexGH{\gamma,\rho,\eta}{\mbSi}$ 
 denotes a zero-mean non-skewed multivariate complex GH distribution
 with a shape parameter $\gamma \in \mathbb{R}$,
 a concentration parameter $\rho > 0$,
 a scaling parameter $\eta > 0$, 
 and a scale matrix $\mbSi \succeq \bm{0}$.
Note that
 the $M$ elements of $\mbz_{ft}$ are mutually dependent
 except for $\mathcal{N}$-FastMNMF, a special case of GH-FastMNMF.
In GH-R1-FastMNMF (GH-FastMNMF with $M=N$ and $\tilde\mbG = \mbI$), 
Eq.~\eqref{eq:p_z_ft_gh} reduces to
\begin{align}
\mbox{GH-R1-FastMNNF:}\
 \mbz_{ft}
 \sim 
 \ComplexGH{\gamma,\rho,\eta}
 {\mathrm{Diag}(\bm\lambda_{ft})},
 \label{eq:p_z_ft_gh_r1}
\end{align}
where $\bm\lambda_{ft} \triangleq [\lambda_{1ft},\cdots,\lambda_{Mft}]^\Tr$
 and the $M$ elements of $\mbz_{ft}$ are assumed to
 have a one-to-one correspondence to $N$ sources ($M=N$).
A reason why the rank-1 version of GH-FastMNMF 
 is called GH-R1-FastMNMF 
 is that the estimated sources are not made independent.
To formulate a generalized hyperbolic extension of ILRMA,
 one can assume a \textit{univariate} complex GH distribution
 for each element of $\mbz_{ft} \triangleq [z_{ft1}, \dots, z_{ftM}]^{\top}$ 
 in exchange for loosing the analytical expression of $\mbx_{ft}$
 (beyond the scope of this paper)
 as follows:
\begin{align}
 z_{ftm}
 \sim 
 \ComplexGH{\gamma,\rho,\eta}
 {\lambda_{mft}}.
 \label{eq:p_z_ft_gh_ilrma}
\end{align}
Note that Eq.~\eqref{eq:p_z_ft_gh_ilrma} is equivalent to Eq.~\eqref{eq:p_z_ft_gh_r1} 
 only for the case of $\mathcal{N}$-FastMNMF,
 because even when an elliptically-contoured multivariate distribution has a \textit{diagonal} scale matrix,
 it cannot generally be factorized into the product of independent dimension-wise univariate distributions.

The LL of the parameters $\mbTh = \{\mbW, \mbH, \mbQ, \tilde\mbG\}$
 is the same in form as Eq.~\eqref{eq:log_likelihood_jdsm},
 where $\log p(\mbz_{ft})$ is given by (see proof in the Appendix)
\begin{align}
\log p(\mbz_{ft})
&\overset{c}{=}
\frac{\gamma - M}{2} 
\log\!
\left(1 + \frac{2}{\rho \eta} \sum_{m=1}^{M} \frac{\tilde{z}_{ftm}}{\tilde{y}_{ftm}}\right)
\nonumber\\
&\quad
+ 
\log \mathcal{K}_{\gamma - M}\!
\left(
\rho
\sqrt{1 + \frac{2}{\rho \eta} \sum_{m=1}^{M} \frac{\tilde{z}_{ftm}}{\tilde{y}_{ftm}}}
\right)
\nonumber\\
&\quad
- 
\sum_{m=1}^{M} \log\tilde{y}_{ftm},
\label{eq:log_likelihood_jdsm_gh}
\end{align}
where $\mathcal{K}_{\zeta}$ denotes
 the modified Bessel function of the second kind 
 with order $\zeta$ \cite{abramowitz1964handbook}.

\subsubsection{Parameter Estimation}

The update rules of the parameters $\mbTh = \{\mbW, \mbH, \mbQ, \tilde\mbG\}$
 are given by Eqs.~\eqref{eq:update_W}--\eqref{eq:norm3},
 where the posterior expectations $\tilde\mbPh$
 are given by Eq.~\eqref{eq:e_log_p_zft}.
As an instance of GSM-FastMNMF,
 GH-FastMNMF is obtained when $\phi_{ft}$ follows a generalized inverse Gaussian (GIG) distribution, denoted $\mathcal{GIG}(\gamma,\rho,\eta)$
 where  $\gamma \in \mathbb{R}$ is a shape parameter,
 $\rho > 0$ is a concentration parameter
and  $\eta > 0$ is a scaling parameter (see Eq.~\eqref{eq:pdf_GIG} for the PDF):
\begin{align}
 \phi_{ft} \sim \mathcal{GIG}(\gamma,\rho,\eta),\label{eq:phi_ft_GIG}
\end{align}
Using Eqs.~\eqref{eq:log_likelihood_jdsm_gh} and \eqref{eq:e_log_p_zft}, we have
\begin{align}
\tilde{\phi}_{ft}^{-1}
&=
\frac{
2(M - \gamma)
}
{
\rho\eta
\left(
1
+ 
\frac{2}{\rho \eta}
\sum_{m=1}^{M}
\frac{\tilde{z}_{ftm}}{\tilde{y}_{ftm}}
\right)
}
\nonumber\\
&\quad
+
\frac{1}
{
\sqrt{\eta}
\sqrt{1 + \frac{2}{\rho \eta} \sum_{m=1}^{M} \frac{\tilde{z}_{ftm}}{\tilde{y}_{ftm}}}
}
\nonumber\\
&\quad\quad
\frac
{\mathcal{K}_{\gamma-M+1}
\left(
\rho
\sqrt{1 + \frac{2}{\rho \eta} \sum_{m=1}^{M} \frac{\tilde{z}_{ftm}}{\tilde{y}_{ftm}}}
\right)}
{
\mathcal{K}_{\gamma-M}
\left(
\rho
\sqrt{1 + \frac{2}{\rho \eta} \sum_{m=1}^{M} \frac{\tilde{z}_{ftm}}{\tilde{y}_{ftm}}}
\right)}
\label{eq:phi-GH}.   
\end{align}
Eq.~\eqref{eq:phi-GH} is already known to appear
 in the estimation of a real univariate GH distribution
 \cite{browne2015mixture, palmer2016algorithm}. 
Interestingly, the same result was found 
 in the estimation of a multivariate isotropic GH distribution.
For mathematical convenience,
 we define an alternative parametrization as follows:
\begin{align}
a \triangleq \frac{\rho}{\eta},
\
b \triangleq \rho\eta.
\label{eq:eta_rho_rep}    
\end{align}
When $\gamma = -\frac{\nu}{2}$, $a = 0$, and $b = \nu$, 
 the general update rules of GSM-FastMNMF given by Eqs.~(\ref{eq:update_W})--(\ref{eq:norm3}) 
 reduce to those of $t$-FastMNMF derived 
 from a lower bound function defined in \cite{kamo2020jointstudent}.\footnote
 {The widely-used multivariate Student's $t$ distribution given by Eq. \eqref{eq:pdf-T}
 is not derived from the multivariate GH distribution given by Eq. \eqref{eq:pdf-GH}.
 In \cite{hu2005calibration}, 
 a multivariate GH distribution with $\gamma = -\nu$, $a = 0$, and $b = \nu$ 
 called a generalized hyperbolic Student's $t$ distribution is used.}
$\mathcal{N}$-FastMNMF is instantiated when $\nu \to \infty$ in $t$-FastMNMF,
 resulting in $\forall (f,t), \tilde{\phi}_{ft}^{-1} = 1$.
Because GH-FastMNMF includes a large variety of distributions, we only consider t-, $\mathcal{N}$
and a new extension based on the normal inverse Gaussian (NIG) distribution more deeply introduced in Section \ref{sec:nig_instance}.
\subsubsection{Source Image Inference}

Using the estimated parameters $\mbTh$,
 we infer the latent source image $\mbx_{nft}$ from the observed mixture $\mbx_{ft}$.
Thanks to the surrogate Gaussian representation
 used in Eqs.~\eqref{eq:p_z_nft_gsm} and \eqref{eq:p_z_ft_gsm},
 the posterior expectation of $\mbx_{nft}$ conditioned by $\phi_{ft}$ 
 can be computed exactly and efficiently with a multichannel Wiener filter as follows:
\begin{align}
 \mathbb{E}\!\left[\mbx_{nft} \mid \mbx_{ft}, \phi_{ft}\right]
 &=
 \mbQ_{f}^{-1} \mathbb{E}\!\left[\mbz_{nft} \mid \mbz_{ft}, \phi_{ft}\right]
 \nonumber\\
 &=
 \mbQ_{f}^{-1} (\phi_{ft}\tilde\mbY_{nft}) \left(\phi_{ft}\tilde\mbY_{ft}\right)^{-1} \mathbf{z}_{ft}
 \nonumber\\
 &=
 \mbQ_{f}^{-1} \tilde\mbY_{nft} \tilde\mbY_{ft}^{-1} \mathbf{z}_{ft},
 \label{eq:marg-Wiener-cond}
\end{align}
where $\phi_{ft}$'s were cancelled out.
We thus have
\begin{align}
 \mathbb{E}\!\left[\mbx_{nft}\mid\mbx_{ft}\right]
 &=
 \mbQ_{f}^{-1} \tilde\mbY_{nft} \tilde\mbY_{ft}^{-1} \mathbf{z}_{ft}.
 \label{eq:marg-Wiener}
\end{align}



\begin{algorithm}[t]
\begin{enumerate}
\setlength{\leftskip}{-0.3cm}
\item \textbf{Input} 
\begin{itemize}
\setlength{\leftskip}{-0.5cm}
\item Multichannel mixture spectrogram~$\mbX$
\end{itemize}
\item \textbf{Configuration} 
\begin{itemize}
\setlength{\leftskip}{-0.5cm}
\item Specify the tail-index parameters
\item[] (except for \mbox{$\mathcal{N}$-FastMNMF})
\begin{align}
\begin{cases}
\nu \ (\mbox{$t$-FastMNMF})
\\
\beta \ (\mbox{GG-FastMNMF})
\\
\rho \ \mbox{and} \ \eta \ (\mbox{NIG-FastMNMF})
\end{cases}
\nonumber
\end{align}
\item Specify the number of bases~$K$
\item Specify the number of iterations~$R$
\end{itemize}
\item \textbf{Initialization}
\begin{itemize}
\setlength{\leftskip}{-0.5cm}
\item Initialize $\mbW$ and $\mbH$ randomly
\item Initialize $\mbQ_f$ to an identity matrix
\item Initialize $\tilde\mbG$ to a circulant matrix given by Eq.~\eqref{eq:circulant_init}
\end{itemize}
\item \textbf{Optimization} For~$r=1\dots R$
\begin{itemize}
\setlength{\leftskip}{-0.5cm}
\item Compute $\tilde{z}_{ftm}$ and $\tilde{y}_{ftm}$
using Eqs.~\eqref{eq:xti_ftm} and \eqref{eq:yti_ftm}, respectively
\item E-step: Compute $\tilde{\phi}_{ft}^{-1} =
 \mathbb{E}_{p(\phi_{ft} \mid \mbz_{ft})}\big[\phi_{ft}^{-1}\big]$ as
\begin{align}
\hspace{-35pt}\tilde{\phi}_{ft}^{-1}
&= 
\begin{cases}
\text{Eq.}~\eqref{eq:e_inv_phi_ft_gaussian}
~(\mbox{$\mathcal{N}$-FastMNMF})
\\
\text{Eq.} ~\eqref{eq:e_inv_phi_ft_student} 
~(\mbox{$t$-FastMNMF})
\\
\text{Eq.}~\eqref{eq:e_inv_phi_ft_gg}
~(\mbox{GG-FastMNMF})
\\
\text{Eq.}~\eqref{eq:phi-NIG}
~(\mbox{NIG-FastMNMF})
\end{cases}\nonumber
\!\!\!\!\!\!\!
\end{align}


\item M-step:
Update $\mbW, \mbH, \tilde{\mbG}$, and $\mbQ$ 
using Eqs. \eqref{eq:update_W}--\eqref{eq:norm3}
\end{itemize}
\item \textbf{Output}
\begin{itemize}
\setlength{\leftskip}{-0.5cm}
\item Source image $\mbX_{n}$ given by Eq.~\eqref{eq:marg-Wiener}
\end{itemize}
\end{enumerate}
\caption{MU-VEM algorithm for GSM-FastMNMF}
\label{alg:GSM-FastMNMF}
\end{algorithm}

\subsection{Normal Inverse Gaussian FastMNMF} 
\label{sec:nig_instance}

As a new variant of GH-FastMNMF with $\gamma = -\frac{1}{2}$,
 we derive NIG-FastMNMF
 based on the normal inverse Gaussian (NIG) distribution. In that case, the Eq.~\eqref{eq:phi-GH} boils down as in \cite{karlis2002type} to:
 \begin{align}
\tilde{\phi}_{ft}^{-1}
&=
\frac{
2(M + \frac{1}{2})
}
{
\rho\eta
\left(
1
+ 
\frac{2}{\rho \eta}
\sum_{m=1}^{M}
\frac{\tilde{z}_{ftm}}{\tilde{y}_{ftm}}
\right)
}
\nonumber\\
&\quad
+
\frac{1}
{
\sqrt{\eta}
\sqrt{1 + \frac{2}{\rho \eta} \sum_{m=1}^{M} \frac{\tilde{z}_{ftm}}{\tilde{y}_{ftm}}}
}
\nonumber\\
&\quad\quad
\frac
{\mathcal{K}_{-M+\frac{1}{2}}
\left(
\rho
\sqrt{1 + \frac{2}{\rho \eta} \sum_{m=1}^{M} \frac{\tilde{z}_{ftm}}{\tilde{y}_{ftm}}}
\right)}
{
\mathcal{K}_{-M-\frac{1}{2}}
\left(
\rho
\sqrt{1 + \frac{2}{\rho \eta} \sum_{m=1}^{M} \frac{\tilde{z}_{ftm}}{\tilde{y}_{ftm}}}
\right)}
\label{eq:phi-NIG}.   
\end{align}
Its constrained version called NIG-R1-FastMNMF
 is obtained when the rank-1 spatial model is used,
 \ie, $M=N$ and $\tilde\mbG = \mbI$.
The NIG distribution is an important sub-class of the GH distribution 
 that is closed under convolution~\cite{podgorski2016convolution}.
Its semi-reproducibility (law linearly stable along with a shape parameter~\cite{hanssen2001normal}) 
 has a high affinity to additivity-aware signal modeling.
For reference,
 the real parts of univariate complex NIG distributions are plotted in Fig.~\ref{fig:GSM-pdf}.

The EM algorithms for $t$-, GG-, and NIG-FastMNMF 
 are obtained as instances of GSM-FastMNMF
 (Algorithm \ref{alg:GSM-FastMNMF}).
\section{Evaluation}\label{sec:evaluation}

This section evaluates the performances of existing and new instances of the proposed GSM-FastMNMF 
 and their rank-1 counterparts
 for a speech enhancement task (Section \ref{sec:speech_enhancement}) 
 and a speech separation task (Section \ref{sec:speech_separation}).
We evaluate the enhanced or the separated speech signals in terms of
the signal-to-distortion ratio (SDR) \cite{vincent2006performance}
and the perceptual evaluation speech quality (PESQ) \cite{itu01pesq}.
 

\subsection{Experimental Conditions}\label{sec:settings}

We compared three existing instances of GSM-FastMNMF
 using the jointly-diagonalizable spatial model (Section \ref{sec:rank_1}),
 \ie, \textbf{${\mathcal{N}}$-FastMNMF} (Section \ref{sec:gaussian_instance}),
 \textbf{$t$-FastMNMF} (Section \ref{sec:t_instance}),
 and \textbf{GG-FastMNMF} (Section \ref{sec:gg_instance})
 with a new instance of GSM-FastMNMF called
 \textbf{NIG-FastMNMF} (Section \ref{sec:nig_instance}),
 where all parameter estimation except for GG-FastMNMF
 are special cases of another instance of GSM-FastMNMF 
 called GH-FastMNMF (Section \ref{sec:GH-FastMNMF}).
Using a determined configuration ($M=N$),
 we also tested the special cases of these methods
 using the rank-1 spatial model (Sections \ref{sec:ex_FastMNMF} \& \ref{sec:GH-FastMNMF}),
 referred to as \textbf{${\mathcal{N}}$-R1-FastMNMF},
 \textbf{$t$-R1-FastMNMF}, \textbf{GG-R1-FastMNMF},
 and \textbf{NIG-R1-FastMNMF}, respectively. 
 Note that \textbf{${\mathcal{N}}$-R1-FastMNMF} is equivalent to \textbf{ILRMA} \cite{kitamura2016determined}.
Heavy-tailed extensions of ILRMA called GG-ILRMA and $t$-ILRMA~\cite{kitamura2018generalized},
 which are different from \textbf{GG-} and \textbf{$t$-R1-FastMNMF} derived in this paper,
 were not considered 
 because they were reported to work no better than ILRMA.
 We also consider \textbf{AuxIVA} \cite{ono2011stable} in the determined case (Fig.~\ref{fig:Init-REVERBC})  and \textbf{OverIVA} \cite{scheibler2019independent} in the overdetermined case (Fig.~\ref{fig:Init-wsj0}). Both IVA versions are computed using a Laplace  model.

We estimated the parameters $\mbTh = \{\mbW, \mbH, \mbQ, \tilde\mbG\}$
 of each method
 for an observed mixture spectrogram $\mbX$
 obtained by applying STFT with a Hann window of 1024 points ($F=513$) and a 75\% overlap
 to a multichannel mixture signal sampled at 16 kHz. 
All elements of the parameters $\mbW$ and $\mbH$ of the NMF-based source model
 were initialized to the absolute values of random samples 
 drawn from a standard Gaussian distribution.
As proposed in \cite{sekiguchi2020fast},
 the parameters $\mbQ$ and $\tilde\mbG$ 
 of the jointly-diagonalizable spatial model
 were initialized as $\mbQ_f \leftarrow \mbI$ and $\tilde\mbG \leftarrow \mbJ$, respectively,
 where $\mbI \in \mathbb{R}_+^{M \times M}$ is an identity matrix
 and $\mbJ \in \mathbb{R}_+^{N \times M}$ is a circulant matrix given by
\begin{align}
\mbJ
=
\left(\begin{array}{ccccccc}
1 & \epsilon & \ldots & \epsilon & 1 & \epsilon & \ldots \\
\epsilon & 1 & \ldots & \epsilon & \epsilon & 1 & \ldots \\
\vdots & \vdots & \ddots & \vdots & \vdots & \vdots & \ddots \\
\epsilon & \epsilon & \ldots & 1 & \epsilon & \epsilon & \ldots
\end{array}\right),
\label{eq:circulant_init}
\end{align}
where $\epsilon$ is set to 
 a small value ($\epsilon = 10^{-2}$ in this paper) for the FastMNMF variants 
 or zero for the R1-FastMNMF variants.

For fair comparison,
 we made two disjoint datasets (validation and test sets)
 in speech enhancement and separation tasks.
In each task, 
 the hyperparameters of each method 
 (\eg, tail indices and the number of NMF bases) 
 were optimized via grid search 
 such that the average SDR on the validation set was maximized. 
For the grid search, we considered
   $\nu\in\{1, 10, 40, 80, 100, 200\}$ for $t$-(R1-)FastMNMF;
    $\beta\in\{1.1, 1.2,$
    $\dots, 1.9\}$ for GG-(R1-)FastMNMF;
    $\rho\in\{1, 5, 10, 15, 20, 30\}$, $\eta\in\{0.5, 1, 2, 3, 5, 10\}$ for NIG-(R1-)FastMNMF,
    and $K \in \{2, 4, 8, 16, 32\}$ for the NMF-based source model. 
The hyperparameters optimized for the validation set and used for the test set
  were listed in Table~\ref{tab:param_se} (speech enhancement) and Table~\ref{tab:param_sp} (speech separation).    
The number of iterations for all methods was set to $300$ 
    because it was enough to optimize FastMNMF, R1-FastMNMF, ILRMA, auxIVA and overIVA until convergence.
The best hyperparameters are then used to evaluate the test set.
Further details on the dataset creation, the hyperparameter optimization, and the best hyperparameter sets
    are described in Section \ref{sec:speech_enhancement} (for speech enhancement task)
    and Section \ref{sec:speech_separation} (for speech separation task).


\subsection{Speech Enhancement with Determined Configurations}
\label{sec:speech_enhancement}

We report a comparative experiment on speech enhancement 
 that aims to extract a \textit{single} speech source from a noisy mixture.
The audio data were taken from the REVERB Challenge dataset~\cite{kinoshita2013reverb},
 where the length of each sample is between 3~[s] and 10~[s].;
Multichannel mixtures ($M \in \{2,5,8\}$) 
 were simulated with a signal-to-noise ratio (SNR) of $0$, $5$, or $10$~dB
 and a reverberation time ($\text{RT}_{60}$) of $250$, $500$, or $700$ ms
 under a \textit{near} or \textit{far} condition 
 that the distance between a microphone array and a speaker
 was $0.5$ or $2.0$~m.
The validation set consists of 
 100 randomly selected mixtures with an SNR of $5$~dB under the near condition.
The test set consists of 
 200 randomly selected mixtures with all conditions.
For fair comparison and 
 the determined nature of the rank-1 spatial model, 
 all methods were used with a determined configuration ($N=M$)
 and the predominant source with the highest average energy 
 was then selected as a target speaker. 
 
\subsubsection{Investigation of Hyperparameters}
\label{sec:hyparam_se}

\begin{table}
\setlength{\tabcolsep}{7pt}
\setlength{\aboverulesep}{1pt}
\setlength{\belowrulesep}{1pt}
\centering
    \caption{Hyperparameters for speech enhancement}
    \label{tab:param_se}
    \vspace{-2mm}
    \begin{tabular}{@{}*{4}{c}@{}}
    \toprule
        \multicolumn{4}{c}{FastMNMF variants}\\
        \cmidrule(lr){1-4}
    $\mathcal{N}$ & $t$ & GG  & NIG  \\
      n/a  &
      $\nu=40$ & 
      $\beta=1.6$ & 
    $(\rho, \eta)=(15, 1)$  
    \\
    $K=4$ & 
    $K=32$ & 
    $K=16$  & 
    $K=8$ 
    \\
        \midrule 
            \multicolumn{4}{c}{R1-FastMNMF variants} \\
            \cmidrule(lr){1-4}
          $\mathcal{N}$ & $t$ & GG  & NIG \\
    n/a & 
    $\nu=40$ & 
    $\beta=1.8$ & 
    $(\rho, \eta)=(10, 1)$ 
    \\
    $K=8$ & 
    $K=4$ & 
    $K=4$  & 
    $K=8$ 
   \\
    \bottomrule             
    \end{tabular}
\end{table}

\begin{table*}[!tp]
\setlength{\tabcolsep}{5.7pt}
\setlength{\aboverulesep}{1pt}
\setlength{\belowrulesep}{1pt}
\begin{centering}
\caption{
The SDRs (mean $\pm$ standard deviation) 
obtained by the eight methods in speech enhancement.
}\label{tab:geo_enhancement_sdr}
\vspace{-2.2mm}
\begin{tabular}{@{}lcc*{8}{r}@{}}
\toprule
\multirow{2}{*}[-2pt]{Dist.} & \multirow{2}{*}[-2pt]{SNR} & \multirow{2}{*}[-2pt]{$M$} & \multicolumn{4}{c}{{FastMNMF variants}} & \multicolumn{4}{c}{{R1-FastMNMF variants}} \\
\cmidrule(lr){4-7} \cmidrule(lr){8-11}
  &  &  &
  \multicolumn{1}{c}{$\mathcal{N}$} & 
  \multicolumn{1}{c}{$t$} &
  \multicolumn{1}{c}{GG} & 
  \multicolumn{1}{c}{NIG} &
  \multicolumn{1}{c}{$\mathcal{N}$} & 
  \multicolumn{1}{c}{$t$} &
  \multicolumn{1}{c}{GG} & 
  \multicolumn{1}{c}{NIG}  \\
\midrule
\multirow{9}{*}[-2pt]{Near} & \multirow{3}{*}{$0$~dB} & 2 &
 $3.6~(\pm 2.2)$ & 
 $3.0~(\pm 1.7)$ &
 $5.1~(\pm 3.7)$ &
 $\mathbf{5.6~(\pm 3.6)}$ &
 $1.8~(\pm 2.0)$ & 
 $3.5~(\pm 5.5)$ &
$1.4~(\pm 4.8)$ &
 $3.2~(\pm 5.0)$ 
 \\
 &  & 5 &
$10.8~(\pm 4.1)$ & 
$8.3~(\pm 2.1)$ & 
$10.9~(\pm 3.5)$ &
$\mathbf{11.9~(\pm 4.1)}$ &
$6.3~(\pm 2.4)$  & 
$6.6~(\pm 4.5)$ &
$6.3~(\pm 4.2)$ &
$7.0~(\pm 3.0)$  
 \\
 
 &  &  8 &
 $12.0~(\pm 4.1)$ &  
 $10.8~(\pm 3.4)$ & 
 $12.8~(\pm 4.3)$ & 
 $\mathbf{13.5~(\pm 4.4)}$ &
 $8.2~(\pm 2.5)$ & 
 $8.8~(\pm 4.1)$ &
  $8.6~(\pm 5.1)$ &
 $8.6~(\pm 2.7)$  
 \\
 
 \cmidrule(lr){2-11}
 & \multirow{3}{*}{$5$~dB} & 2 &
$9.7~(\pm 4.4)$ &  
$8.1~(\pm 2.8)$ &
$10.1~(\pm 3.8)$ &
$\mathbf{10.3~(\pm 3.8)}$ &
$6.2~(\pm 1.7)$ & 
$7.1~(\pm 5.1)$ &  
$6.3~(\pm 4.9)$ & 
$6.5~(\pm 2.7)$ 
 \\
 &  &  5 &  
$13.2~(\pm 3.4)$ & 
$12.1~(\pm 1.8)$ & 
$14.1~(\pm 2.9)$ &
$\mathbf{14.7~(\pm 3.4)}$ &
$10.0~(\pm 1.7)$ & 
$11.5~(\pm 3.3)$ &
$11.9~(\pm 3.9)$ &
$11.4~(\pm 3.0)$ 
 \\
 &  &  8 & 
 $14.4~(\pm 3.2)$ &  
 $14.1~(\pm 2.7)$ & 
 $15.7~(\pm 3.2)$ &
 $\mathbf{16.0~(\pm 3.6)}$ &
  $11.7~(\pm 2.3)$ & 
   $11.1~(\pm 3.5)$ & 
    $13.8~(\pm 3.1)$ &
 $13.3~(\pm 2.6)$  
 \\
 
 \cmidrule(lr){2-11}
 & \multirow{3}{*}{$10$~dB} & 2 &
$12.2~(\pm 3.6)$ & 
$11.7~(\pm 2.9)$ &
$\mathbf{13.4~(\pm 3.4)}$ &
$13.4~(\pm 3.5)$ &
$9.5~(\pm 2.0)$ & 
$10.8~(\pm 3.5)$ &
$11.2~(\pm 3.7)$ &
$11.7~(\pm 2.2)$  

 \\
 &  &  5 &  
$14.5~(\pm 3.2)$ &
$14.7~(\pm 2.0)$ &
$16.2~(\pm 2.9)$ &
$\mathbf{16.3~(\pm 3.3)}$ &
$12.8~(\pm 1.8)$ & 
$12.6~(\pm 3.0)$ & 
$13.6~(\pm 3.5)$ & 
$13.8~(\pm 3.6)$ 
 \\
 &  &  8 &
  $15.2~(\pm 3.0)$ & 
   $15.7~(\pm 2.4)$ & 
    $17.0~(\pm 2.9)$ & 
 $\mathbf{17.6~(\pm 3.1)}$ &
  $14.6~(\pm 2.2)$ & 
   $12.7~(\pm 2.9)$ &
   $14.3~(\pm 3.2)$ &
 $14.4~(\pm 3.5)$ 

 \\
 \midrule 
\multirow{9}{*}[-2pt]{Far} & \multirow{3}{*}{$0$~dB} & 2 &
$2.1~(\pm 4.8)$ &
$0.5~(\pm 1.8)$ &
$1.5~(\pm 2.2)$ & 
$\mathbf{2.2~(\pm 2.8)}$ &
$-0.8~(\pm 2.4)$ & 
$0.7~(\pm 4.8)$ &
$0.6~(\pm 4.7)$ &
$1.1~(\pm 2.1)$ 
\\
 &  & 5 &
 $5.4~(\pm 4.6)$ &
 $4.6~(\pm 2.8)$ & 
 $6.5~(\pm 4.2)$ &
$\mathbf{7.2~(\pm 4.5)}$ &
$2.7~(\pm 2.7)$ & 
$3.7~(\pm 3.1)$ &
$3.3~(\pm 5.7)$ & 
$3.7~(\pm 2.7)$  
\\
 &  & 8 &
 $6.3~(\pm 4.0)$ & 
 $6.1~(\pm 3.4)$ & 
 $7.7~(\pm 4.0)$ &
$\mathbf{8.2~(\pm 3.9)}$ &
 $4.1~(\pm 3.3)$ & 
 $5.2~(\pm 3.2)$ &
 $5.7~(\pm 4.1)$ & 
$5.9~(\pm 2.8)$  
\\
\cmidrule(lr){2-11}
& \multirow{3}{*}{$5$~dB} & 2 &
$4.9~(\pm 4.4)$ &
$3.7~(\pm 2.2)$ &
$5.0~(\pm 3.0)$ &
$\mathbf{5.4~(\pm 3.3)}$ &
$2.7~(\pm 2.4)$ & 
$3.4~(\pm 3.1)$ & 
 $3.7~(\pm 3.0)$ &
$3.6~(\pm 2.7)$ 
\\
&  & 5 &
$6.8~(\pm 4.3)$ &
$7.2~(\pm 3.4)$ &
$8.3~(\pm 4.2)$ &
$\mathbf{8.5~(\pm 4.3)}$ &
 $5.6~(\pm 3.4)$ & 
 $4.2~(\pm 4.2)$ &
 $4.9~(\pm 3.6)$ &
$5.2~(\pm 2.9)$  
 \\
 &  & 8 &
 $8.1~(\pm 3.5)$ &
 $8.4~(\pm 3.2)$ &
 $9.5~(\pm 3.6)$ &
$\mathbf{9.6~(\pm 3.6)}$ &
$6.2~(\pm 4.0)$ & 
$7.9~(\pm 3.2)$ & 
$7.7~(\pm 3.5)$ &
$8.3~(\pm 3.3)$ 
\\

\cmidrule(lr){2-11}
& \multirow{3}{*}{$10$~dB} & 2 &
$5.9~(\pm 4.2)$ & 
$5.9~(\pm 2.9)$ & 
$7.1~(\pm 3.6)$ &
$\mathbf{7.2~(\pm 3.5)}$ &
$4.8~(\pm 3.3)$ & 
$5.2~(\pm 4.1)$ &
$5.2~(\pm 4.4)$ &
$5.6~(\pm 3.1)$ 
\\
&  &  5 & 
$7.7~(\pm 4.4)$ & 
$8.7~(\pm 3.7)$ &
$\mathbf{9.7~(\pm 4.2)}$ & 
$9.6~(\pm 4.2)$ &
 $7.3~(\pm 4.1)$ & 
 $8.0~(\pm 5.6)$ &
 $7.5~(\pm 4.5)$ &
$8.2~(\pm 3.1)$  
\\
&  &  8 &
$8.9~(\pm 3.5)$ &
$9.9~(\pm 3.3)$ & 
$\mathbf{10.7~(\pm 3.6)}$ &
$10.6~(\pm 3.6)$ &
$8.3~(\pm 4.6)$ & 
$9.0~(\pm 4.2)$ &
$9.6~(\pm 4.4)$ &
$9.2~(\pm 3.2)$ 
\\
\bottomrule
\end{tabular}
\par\end{centering}
\end{table*}

\begin{table*}[!tp]
\setlength{\tabcolsep}{7.5pt}
\setlength{\aboverulesep}{1pt}
\setlength{\belowrulesep}{1pt}
\begin{centering}
\caption{
The PESQs (mean $\pm$ standard deviation) 
obtained by the eight methods in speech enhancement.}
\label{tab:geo_enhancement_pesq}
\vspace{-2.2mm}
\begin{tabular}{@{}lcc*{8}{r}@{}}
\toprule
\multirow{2}{*}[-2pt]{Dist.} & \multirow{2}{*}[-2pt]{SNR} & \multirow{2}{*}[-2pt]{$M$} & \multicolumn{4}{c}{{FastMNMF variants}} & \multicolumn{3}{c}{{R1-FastMNMF variants}} \\
\cmidrule(lr){4-7} \cmidrule(lr){8-11}
  &  &  &
  \multicolumn{1}{c}{$\mathcal{N}$} &
  \multicolumn{1}{c}{$t$} & 
   \multicolumn{1}{c}{GG} & 
  \multicolumn{1}{c}{NIG} &

  \multicolumn{1}{c}{$\mathcal{N}$} &
  \multicolumn{1}{c}{$t$} &
  \multicolumn{1}{c}{GG} &
  \multicolumn{1}{c}{NIG} 
  \\
\midrule

\multirow{9}{*}[-2pt]{Near} & \multirow{3}{*}{$0$~dB} & 2 &
$1.8~(\pm 0.6)$ & 
$1.9~(\pm 0.6)$ & 
$\mathbf{2.0~(\pm 0.7)}$ & 
$1.9~(\pm 0.6)$ & 
$1.7~(\pm 0.6)$ & 
$1.7~(\pm 0.6)$ & 
$1.8~(\pm 0.6)$ & 
$1.7~(\pm 0.6)$ \\   
&  & 5 &
$2.3~(\pm 0.7)$ & 
$\mathbf{2.4~(\pm 0.7)}$  & 
$\mathbf{2.4~(\pm 0.7)}$  & 
$\mathbf{2.4~(\pm 0.7)}$ &
$2.1~(\pm 0.7)$ &
$2.0~(\pm 0.7)$ &
$2.0~(\pm 0.7)$ &
$1.9~(\pm 0.7)$  \\  

&  & 8 &
$2.4~(\pm 0.7)$ &  
$2.5~(\pm 0.7)$ &  
$\mathbf{2.6~(\pm 0.8)}$ &  
$\mathbf{2.6~(\pm 0.8)}$ & 
$2.2~(\pm 0.8)$  & 
$2.3~(\pm 0.7)$ &  
$2.3~(\pm 0.7)$ &  
$2.1~(\pm 0.7)$   
\\
\cmidrule(lr){2-11}
& \multirow{3}{*}{$5$~dB} & 2 &
$2.1~(\pm 0.7)$ & 
$2.2~(\pm 0.6)$ & 
$\mathbf{2.2~(\pm 0.7)}$ & 
$\mathbf{2.2~(\pm 0.7)}$  & 
$2.0~(\pm 0.6)$ & 
$2.1~(\pm 0.7)$ & 
$2.0~(\pm 0.7)$ & 
$1.9~(\pm 0.7)$  
\\
&  &  5 & 
$2.6~(\pm 0.6)$ & 
$\mathbf{2.7~(\pm 0.6)}$ & 
$\mathbf{2.7~(\pm 0.6)}$ &  
$2.7~(\pm 0.7)$ & 
$2.4~(\pm 0.7)$ & 
$2.4~(\pm 0.8)$ & 
$2.5~(\pm 0.8)$ & 
$2.0~(\pm 0.9)$ 
\\
&  &  8 & 
$2.8~(\pm 0.6)$ & 
$2.8~(\pm 0.7)$ & 
$\mathbf{2.9~(\pm 0.7)}$ & 
$\mathbf{2.9~(\pm 0.7)}$ & 
$2.5~(\pm 0.7)$ &
$2.6~(\pm 0.8)$ & 
$2.6~(\pm 0.8)$ & 
$2.2~(\pm 0.8)$  
\\
\cmidrule(lr){2-11}
& \multirow{3}{*}{$10$~dB} & 2 &
$2.3~(\pm 0.6)$ &  
$2.4~(\pm 0.6)$ &  
$\mathbf{2.5~(\pm 0.6)}$ &  
$\mathbf{2.5~(\pm 0.6)}$ &  
$2.2~(\pm 0.7)$ & 
$2.3~(\pm 0.8)$ &  
$2.2~(\pm 0.8)$ &  
$2.0~(\pm 0.8)$   
\\

&  &  5 & 
$2.8~(\pm 0.5)$ & 
$\mathbf{3.0~(\pm 0.5)}$  & 
$\mathbf{3.0~(\pm 0.5)}$ & 
$\mathbf{3.0~(\pm 0.5)}$ & 
$2.7~(\pm 0.6)$ & 
$2.7~(\pm 0.9)$ & 
$2.7~(\pm 0.9)$ & 
$2.4~(\pm 0.9)$  
\\

&  &  8 & 
$3.0~(\pm 0.5)$ & 
$3.1~(\pm 0.5)$ & 
$\mathbf{3.2~(\pm 0.5)}$ & 
$\mathbf{3.2~(\pm 0.5)}$ & 
$2.8~(\pm 0.6)$ & 
$2.9~(\pm 0.9)$ & 
$2.9~(\pm 0.9)$ & 
$2.7~(\pm 0.9)$  
\\
\midrule 
\multirow{9}{*}[-2pt]{Far} & \multirow{3}{*}{$0$~dB} & 2 &
$1.6~(\pm 0.4)$ & 
$\mathbf{1.7~(\pm 0.4)}$ &  
$\mathbf{1.7~(\pm 0.4)}$ & 
$\mathbf{1.7~(\pm 0.4)}$ & 

$1.5~(\pm 0.4)$ & 
$1.5~(\pm 0.4)$ & 
$1.5~(\pm 0.4)$ & 
$1.5~(\pm 0.4)$   
\\
&  & 5 &
$1.9~(\pm 0.5)$ & 
$2.0~(\pm 0.5)$ &  
$\mathbf{2.1~(\pm 0.5)}$ & 
$2.1~(\pm 0.6)$ & 
$1.8~(\pm 0.5)$ & 
$1.8~(\pm 0.5)$ & 
$1.9~(\pm 0.5)$ & 
$1.8~(\pm 0.5)$  
\\

&  & 8 &
$2.1~(\pm 0.6)$ & 
$2.2~(\pm 0.6)$ & 
$2.2~(\pm 0.6)$ & 
$\mathbf{2.3~(\pm 0.7)}$ & 
$1.9~(\pm 0.6)$ & 
$1.9~(\pm 0.5)$ & 
$1.9~(\pm 0.5)$ & 
$1.9~(\pm 0.5)$  
\\

\cmidrule(lr){2-11}
& \multirow{3}{*}{$5$~dB} & 2 &
$1.7~(\pm 0.4)$ & 
$\mathbf{1.9~(\pm 0.4)}$  & 
$\mathbf{1.9~(\pm 0.4)}$ & 
$\mathbf{1.9~(\pm 0.4)}$ & 
$1.7~(\pm 0.4)$ & 
$1.7~(\pm 0.5)$ & 
$1.7~(\pm 0.5)$ &  
$1.6~(\pm 0.5)$  

\\
&  &  5 & 
$2.1~(\pm 0.5)$ & 
$\mathbf{2.2~(\pm 0.4)}$ &  
$2.2~(\pm 0.5)$ & 
$\mathbf{2.2~(\pm 0.4)}$ &  
$2.0~(\pm 0.5)$ &
$2.0~(\pm 0.6)$ &  
$2.1~(\pm 0.5)$ &  
$1.9~(\pm 0.6)$ 
\\
&  &  8 & 
$2.1~(\pm 0.5)$ & 
$2.3~(\pm 0.5)$ & 
$\mathbf{2.4~(\pm 0.6)}$ & 
$2.3~(\pm 0.5)$ & 
$2.1~(\pm 0.6)$ & 
$2.2~(\pm 0.6)$ & 
$2.2~(\pm 0.6)$ & 
$2.0~(\pm 0.6)$  
\\

\cmidrule(lr){2-11}
& \multirow{3}{*}{$10$~dB} & 2 &
$1.8~(\pm 0.4)$ & 
$\mathbf{2.0~(\pm 0.4)}$ & 
$\mathbf{2.0~(\pm 0.4)}$ & 
$\mathbf{2.0~(\pm 0.4)}$ & 
$1.8~(\pm 0.4)$ & 
$1.8~(\pm 0.5)$ & 
$1.8~(\pm 0.5)$ & 
$1.7~(\pm 0.5)$  

\\
&  &  5 & 
$2.1~(\pm 0.4)$ & 
$\mathbf{2.3~(\pm 0.4)}$ & 
$\mathbf{2.3~(\pm 0.4)}$ &  
$\mathbf{2.3~(\pm 0.4)}$ &  
$2.1~(\pm 0.5)$ & 
$2.2~(\pm 0.6)$ & 
$2.1~(\pm 0.6)$ & 
$2.0~(\pm 0.6)$  
\\
&  &  8 & 
$2.3~(\pm 0.5)$ & 
$\mathbf{2.5~(\pm 0.5)}$ & 
$\mathbf{2.5~(\pm 0.5)}$ & 
$\mathbf{2.5~(\pm 0.5)}$ &  
$2.3~(\pm 0.5)$ & 
$2.4~(\pm 0.6)$ & 
$2.3~(\pm 0.6)$ & 
$2.4~(\pm 0.6)$  
\\
\bottomrule
\end{tabular}
\par\end{centering}
\end{table*}
  
\begin{table*}
\setlength{\tabcolsep}{5.7pt}
\setlength{\aboverulesep}{1pt}
\setlength{\belowrulesep}{1pt}
\begin{centering}
\caption{
The SDRs (mean $\pm$ standard deviation) 
obtained by the eight methods in speech enhancement for $M=8$ and various $RT_{60}$.
}\label{tab:geo_enhancement_sdr_RT60}
\vspace{-2mm}
\begin{tabular}{@{}lcc*{8}{r}@{}}
\toprule
\multirow{2}{*}[-2pt]{Dist.} & \multirow{2}{*}[-2pt]{SNR} & \multirow{2}{*}[-2pt]{$RT_{60}$ [s]} & \multicolumn{4}{c}{{FastMNMF variants}} & \multicolumn{4}{c}{{R1-FastMNMF variants}} \\
\cmidrule(lr){4-7} \cmidrule(lr){8-11}
  &  &  &
  \multicolumn{1}{c}{$\mathcal{N}$} & 
  \multicolumn{1}{c}{$t$} &
  \multicolumn{1}{c}{GG} & 
  \multicolumn{1}{c}{NIG} &
  \multicolumn{1}{c}{$\mathcal{N}$} & 
  \multicolumn{1}{c}{$t$} &
  \multicolumn{1}{c}{GG} & 
  \multicolumn{1}{c}{NIG}  \\
\midrule
\multirow{9}{*}[-2pt]{Near} & \multirow{3}{*}{$0$~dB} & 0.25 &
$14.3~(\pm 4.2)$ &
$13.8~(\pm 3.2)$ &
$15.9~(\pm 4.6)$ &
$\mathbf{16.5~(\pm 4.3)}$ &
$11.9~(\pm 2.9)$ &
$12.2~(\pm 3.3)$ & 
$11.6~(\pm 2.6)$ & 
$12.6~(\pm 2.8)$
 \\
 &  & 0.50 &
$11.1~(\pm 4.1)$ &
$11.0~(\pm 3.8)$ &
$12.0~(\pm 4.4)$ &
$\mathbf{12.9~(\pm 4.2)}$ & 
$7.1~(\pm 2.1)$ & 
$7.9~(\pm 2.7)$ & 
$7.6~(\pm 3.2)$ & 
$8.3~(\pm 2.3)$
 \\
 
 &  &  0.70 &
$10.7~(\pm 3.8)$ &
$7.7~(\pm 2.9)$ &
$10.4~(\pm 3.9)$ & 
$\mathbf{11.1~(\pm 4.5)}$ & 
$5.6~(\pm 2.4)$ &  
$7.1~(\pm 3.2)$ &  
$6.6~(\pm 3.9)$ &  
$7.7~(\pm 3.0)$
 \\
 
 \cmidrule(lr){2-11}
 & \multirow{3}{*}{$5$~dB} & 0.25 &
$18.0~(\pm 3.2)$ & 
$16.7~(\pm 2.6)$ & 
$18.1~(\pm 3.3)$ &  
$\mathbf{18.4~(\pm 3.3)}$ & 
$14.9~(\pm 2.5)$ &  
$15.2~(\pm 2.7)$ &  
$14.2~(\pm 3.6)$ & 
$15.4~(\pm 2.8)$
 \\
 &  &  0.50 &  
$13.1~(\pm 3.0)$ &
$13.3~(\pm 2.4)$ &
$14.8~(\pm 3.1)$ &
$\mathbf{15.8~(\pm 4.2)}$ &
$11.2~(\pm 2.5)$ & 
$11.2~(\pm 3.0)$ & 
$11.7~(\pm 2.9)$ &
$13.0~(\pm 2.3)$ 
 \\
 &  &  0.70 & 
$12.3~(\pm 3.4)$ & 
$12.2~(\pm 3.0)$ &
$\mathbf{14.2~(\pm 3.2)}$ &
$13.8~(\pm 3.2)$ & 
$9.1~(\pm 2.0)$ & 
$11.0~(\pm 2.8)$ & 
$10.7~(\pm 1.9)$ & 
$11.6~(\pm 2.6)$
\\
 
 \cmidrule(lr){2-11}
 & \multirow{3}{*}{$10$~dB} & 0.25 &
$19.5~(\pm 3.8)$ &
$18.7~(\pm 2.3)$ & 
$19.1~(\pm 3.3)$ &
$\mathbf{19.9~(\pm 2.8)}$ &
$18.0~(\pm 2.1)$ &  
$17.2~(\pm 3.1)$ &  
$16.2~(\pm 2.7)$ &  
$18.5~(\pm 3.1)$
 \\
 &  &  0.50 &  
$13.9~(\pm 2.7)$ &
$14.3~(\pm 2.3)$ &
$16.8~(\pm 2.8)$ &
$\mathbf{17.1~(\pm 3.1)}$ &  
$14.2~(\pm 2.5)$ & 
$12.1~(\pm 6.9)$ &
$13.5~(\pm 2.7)$ & 
$13.8~(\pm 4.6)$\\
 
 &  &  0.70 &
$12.3~(\pm 1.9)$ & 
$14.2~(\pm 2.4)$ &
$15.0~(\pm 2.5)$ & 
$\mathbf{15.8~(\pm 3.4)}$ & 
$11.6~(\pm 2.1)$ &
$11.2~(\pm 2.6)$ & 
$10.6~(\pm 5.7)$ &  
$13.6~(\pm 2.3)$
 \\
 \midrule 
\multirow{9}{*}[-2pt]{Far} & \multirow{3}{*}{$0$~dB} & 0.25 &
$9.9~(\pm 3.4)$ & 
$7.8~(\pm 3.5)$ & 
$10.2~(\pm 4.4)$ & 
$\mathbf{10.9~(\pm 3.8)}$ &  
$8.3~(\pm 3.4)$ &  
$8.0~(\pm 3.4)$ & 
$7.0~(\pm 3.1)$ &  
$9.0~(\pm 2.4)$
\\
 &  & 0.50 &
$4.9~(\pm 3.9)$ & 
$5.9~(\pm 3.6)$ & 
$6.8~(\pm 3.6)$ &
$\mathbf{7.0~(\pm 4.1)}$ & 
$3.9~(\pm 2.1)$ & 
$3.7~(\pm 2.7)$ & 
$4.5~(\pm 3.3)$ &
$5.0~(\pm 2.9)$
\\
 &  & 0.70 &
$4.0~(\pm 4.7)$ &
$4.6~(\pm 3.0)$ &
$6.2~(\pm 4.3)$ & 
$\mathbf{6.6~(\pm 3.8)}$ &  
$0.9~(\pm 2.4)$ &  
$3.5~(\pm 2.7)$ &  
$3.2~(\pm 3.6)$ &  
$4.2~(\pm 3.0)$
\\
\cmidrule(lr){2-11}
& \multirow{3}{*}{$5$~dB} & 0.25 &
 $11.7~(\pm 4.1)$ & 
 $11.3~(\pm 3.3)$ & 
 $11.7~(\pm 3.5)$ & 
 $\mathbf{12.2~(\pm 3.9)}$ & 
 $10.0~(\pm 2.8)$ & 
 $10.2~(\pm 2.8)$ & 
 $9.7~(\pm 4.9)$ &  
 $9.5~(\pm 3.0)$
\\
&  & 0.50 &
$6.8~(\pm 3.1)$ & 
$7.0~(\pm 2.7)$ & 
$\mathbf{8.7~(\pm 3.5)}$ & 
$8.5~(\pm 3.5)$ & 
$5.8~(\pm 3.5)$ &
$6.4~(\pm 3.2)$ & 
$6.3~(\pm 2.4)$ &  
$8.0~(\pm 4.1)$
 \\
 &  & 0.70 &
$5.8~(\pm 3.3)$ & 
$7.0~(\pm 3.6)$ & 
$8.2~(\pm 3.8)$ &
$\mathbf{8.3~(\pm 3.4)}$ &  
$3.9~(\pm 3.9)$ &  
$4.6~(\pm 3.2)$ & 
$5.5~(\pm 3.2)$ & 
$7.3~(\pm 2.6)$ 
\\

\cmidrule(lr){2-11}
& \multirow{3}{*}{$10$~dB} & 0.25 &
$13.0~(\pm 3.0)$ &
$12.0~(\pm 3.2)$ & 
$\mathbf{13.8~(\pm 3.8)}$ &
$13.7~(\pm 3.8)$ &  
$12.1~(\pm 4.1)$ & 
$11.8~(\pm 3.1)$ & 
$9.6~(\pm 3.7)$ & 
$11.2~(\pm 3.9)$
\\
&  &  0.50 & 
$8.7~(\pm 3.8)$ &
$\mathbf{10.1~(\pm 2.8)}$ &  
$9.4~(\pm 2.9)$ &
$9.8~(\pm 4.0)$ &
$9.1~(\pm 3.0)$ &  
$6.8~(\pm 2.3)$ & 
$7.5~(\pm 3.2)$ &
$8.5~(\pm 3.3)$
\\
&  &  0.70 &
$5.0~(\pm 3.6)$ &
$7.6~(\pm 3.6)$ &
$\mathbf{9.0~(\pm 3.8)}$ & 
$8.3~(\pm 2.7)$ & 
$4.9~(\pm 4.1)$ & 
$4.7~(\pm 4.9)$ & 
$7.3~(\pm 4.0)$ & 
$7.9~(\pm 2.6)$
\\
\bottomrule
\end{tabular}
\par\end{centering}
\end{table*}

\begin{table}
\setlength{\tabcolsep}{7.5pt}
\setlength{\aboverulesep}{1pt}
\setlength{\belowrulesep}{1pt}
\begin{centering}
\caption{ Statistical significance (  "***" denotes high ($p< 0.001$), "**" good ($p<0.01$), "*" marginal ($p<0.05$) and "n.s." non significant ($p>=0.05$) p-value)
 for a non-parametric Wilcoxon tested on the NIG-(R1)FastMNMF  SDR scores obtained in Section \ref{sec:speech_enhancement}}\label{tab:stat_REVERBC}
\begin{tabular}{@{}lcc*{7}{r}@{}}
\toprule
\multirow{2}{*}[-2pt]{$K$}  & \multicolumn{3}{c}{{FastMNMF variants}} & \multicolumn{3}{c}{{R1-FastMNMF variants}} \\
\cmidrule(lr){2-4}\cmidrule(lr){5-7}
  & \multicolumn{1}{c}{$\mathcal{N}$} & \multicolumn{1}{c}{$t$} & \multicolumn{1}{c}{GG}&
  \multicolumn{1}{c}{$\mathcal{N}$} & \multicolumn{1}{c}{$t$} & \multicolumn{1}{c}{GG}
   \\
\midrule
2 &
** &
**  &
* &
* &
*** &
n.s.
\\
 4 &
* &
**  &
*** &
* &
** &
*

\\

 8 &
*** &
**  &
* &
** &
** &
**
\\
 16 &
** &
**  &
* &
n.s. &
** &
*
\\

 32 &
** &
*  &
* &
** &
*** &
*
\\
\bottomrule
\end{tabular}
\par\end{centering}
\end{table}

\begin{table}
\setlength{\tabcolsep}{7pt}
\setlength{\aboverulesep}{1pt}
\setlength{\belowrulesep}{1pt}
\centering
    \caption{Per-iteration times [s] with $K=16, N=M=8$ (GPU/CPU)}
    \label{tab:elapsed_time}
    \vspace{-2mm}
    \begin{tabular}{@{}*{4}{c}@{}}
    \toprule
        \multicolumn{4}{c}{FastMNMF variants}\\
        \cmidrule(lr){1-4}
         $\mathcal{N}$ & $t$ & GG & NIG \\
        $\mathbf{0.012}/0.536$ & 
        $0.012/\mathbf{0.535}$ & 
        $0.012/0.537$ &  
        $0.025/0.655$  
        \\
        \midrule 
        \multicolumn{4}{c}{R1-FastMNMF variants}\\
        \cmidrule(lr){1-4}
        $\mathcal{N}$ & $t$ & GG & NIG  \\
        $0.006/0.169$ & 
        $\mathbf{0.006}/\mathbf{0.137}$ & 
        $0.006/0.171$ &  
        $0.021/0.232$  
        \\
    \bottomrule             
    \end{tabular}
\end{table}
\begin{figure}[t]
\centering
\includegraphics[width=.9\columnwidth]{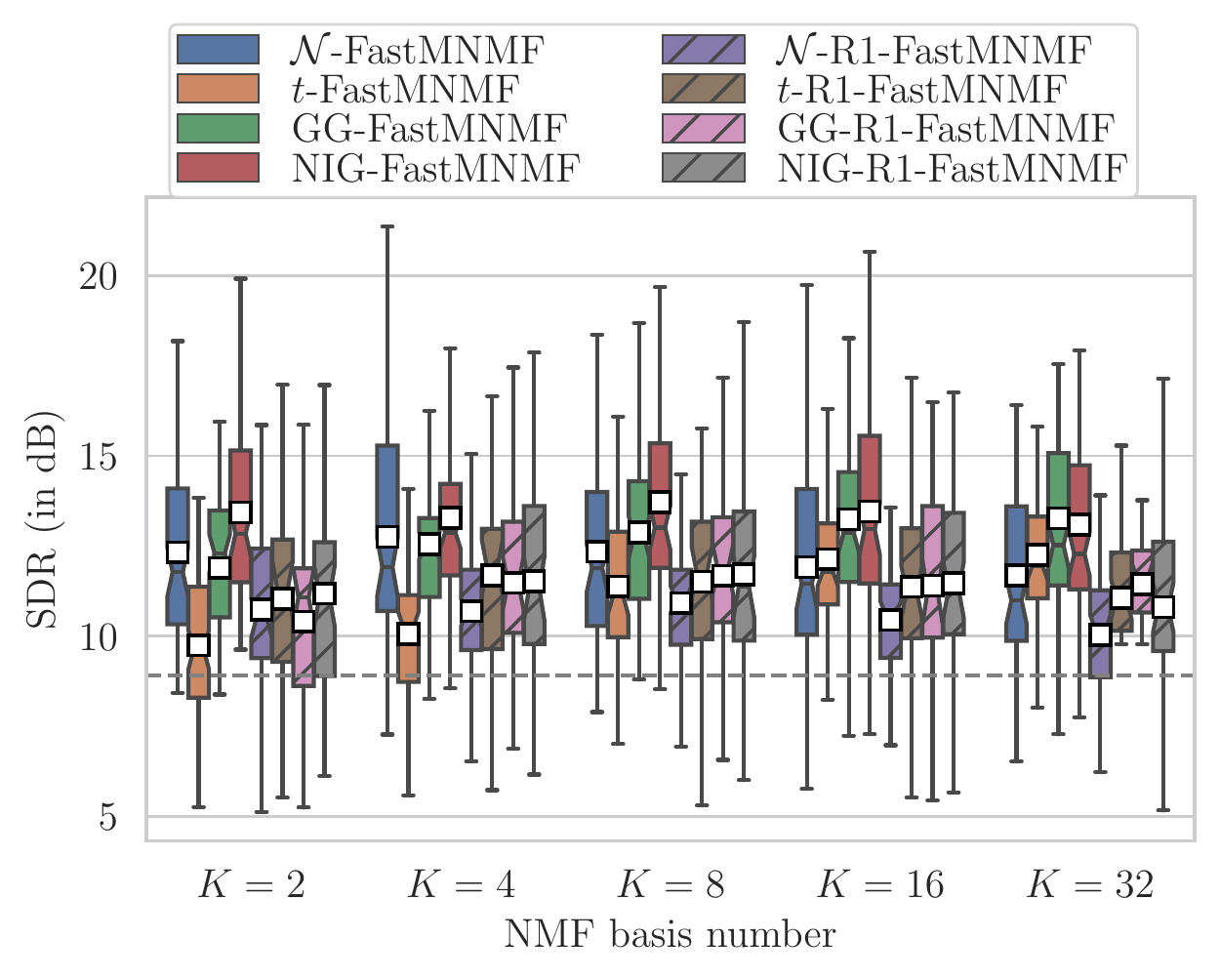}
\vspace{-2.8mm}
\caption{
The SDRs obtained by $\mathcal{N}$-(R1-)FastMNMF, 
$t$-(R1-)FastMNMF, GG-(R1-)FastMNMF, and NIG-(R1-)FastMNMF with $K\in\{2,4,8,16,32\}$
in speech enhancement. 
White squares and notches indicate the means and the 95\% confidence intervals, respectively.
The dashed grey line represents the median SDR results obtained by AuxIVA.}
\label{fig:Init-REVERBC}
\end{figure}

The optimal parameters for the speech enhancement task
    based on the grid search parameter optimization (see Section \ref{sec:settings})
    are listed in Table~\ref{tab:param_se}.
%
%
Fig.~\ref{fig:Init-REVERBC} shows the SDRs on the validation set
 obtained by the eight methods with $M=N=8$ and $K\in\{2,4,8,16,32\}$ while Table~\ref{tab:stat_REVERBC} reports statistical significance results based on Wilcoxon tests \cite{wilcoxon1992individual} between NIG(R1)-FastMNMF scores in Fig.~\ref{fig:Init-REVERBC} and other extensions.
NIG-FastMNMF tended to outperform the other methods
 and attained the best median  and mean SDRs when $K=8$ with a statistical significance of $ p \approx 0.011$ in average,
 whereas GG-FastMNMF attained the best SDR when $K=16$.
 We found the 95\% confidence interval and interquartile range of NIG-FastMNMF 
 was wider than those of $\mathcal{N}$-FastMNMF. 
This could be explained by the numerical instability 
 of approximating the ratio of the modified Bessel functions in Eq.~\eqref{eq:phi-GH}.
 
In contrast, $t$-FastMNMF with a larger $K$ gave a better SDR
 and $\mathcal{N}$-FastMNMF with $K=4$ achieved the best SDR.
As noticed in \cite{kitamura2016determined, kitamura2018generalized},
 we observed that the rank-1 variants with a smaller $K$ tended to work better.
Among the R1-FastMNMF variants, 
 $t$-R1-FastMNMF with $K=4$ achieved the best SDR.

\subsubsection{Investigation of Performances}
\label{sec:se_results}


Tables~\ref{tab:geo_enhancement_sdr} and \ref{tab:geo_enhancement_pesq}
 respectively show the SDRs and PESQs on the test set
 obtained by the eight methods with the optimized hyperparameters.
For any method under any condition,
 the use of more microphones resulted in a better SDR and PESQ.

In terms of the SDR,
 NIG-FastMNMF worked best on average under most conditions
 and outperformed the other methods by a larger margin 
 under a more adverse condition (\eg, SNR of $0$~dB).
In terms of the PESQ,
 GG-FastMNMF worked best on average when $M \in \{2,5\}$,
 whereas NIG-FastMNMF generally worked best when $M=8$.
Since the modified Bessel function in Eq.~\eqref{eq:phi-GH} 
 is hard to compute with a high degree of precision,
 the perceptual quality might have been degraded by some artifacts.

Table~\ref{tab:geo_enhancement_sdr_RT60} shows the SDRs on the test set obtained when $M=8$. 
As a whole, 
 heavy-tailed extensions worked more accurately as the $RT_{60}$ decreases. 
In most cases, NIG-FastMNMF was slightly better than the other variants 
 except for the far setting with an SNR of $10$~dB.

In terms of the SDR, 
 the heavy-tailed R1-FastMNMF variants worked comparably on average when $M=8$,
 albeit NIG-R1-FastMNMF achieved the lowest standard deviation.
This indicates the robustness of NIG-R1-FastMNMF against various SNR and distance conditions.
A similar result was nevertheless not observed in terms of the PESQ. 

Overall, the proposed NIG-FastMNMF 
 is considered to be the most reasonable choice
 in a real scenario in terms of the SDR and PESQ. 
Table~\ref{tab:elapsed_time} lists the average elapsed times
 of the eight methods with $K=16$
 on a GPU (NVIDIA® TITAN RTX™)
 or CPU (Intel® Xeon® W-2145).
The relatively heavier computation of the NIG variants 
 were originated from the modified Bessel function 
 used in Eq.~\eqref{eq:phi-GH}.
This issue could be solved
 with a more efficient library than \textit{scipy}\cite{scipy}.

\subsection{Speech Separation with (Over)determined Configurations}
\label{sec:speech_separation}

We report a comparative experiment on speech separation
 that aims to separate \textit{multiple} speech sources 
 from an echoic mixture in the overdetermined case $M>N$.
The audio data were taken 
 from the WSJ0-mix reverberant dataset \cite{wang2018combining, wang2018multi}
 where each sample is between 3~[s] and 8~[s] long
 and includes $N\in\left\{2,3\right\}$ speakers
 with an $\text{RT}_{60}$ randomly ranging from $200$ [ms] to $700$ [ms].
The \textit{validation set} consists of $100$ utterances
 and the \textit{test set} consists of $200$ utterances.
For fair comparison, 
 $\mathcal{N}$-, $t$-, GG-, and NIG-FastMNMF
 were tested with both determined ($N=M\in\left\{2,3\right\}$)
 and overdetermined ($M\in\left\{5,8\right\}$) configurations,
 where $N$ sources with the highest average energies 
 were selected as target speakers.

\subsubsection{Investigation of Hyperparameters}
\label{sec:hyparam_ss}

\begin{figure}[!ht]
\centering
\includegraphics[width=0.85\columnwidth]{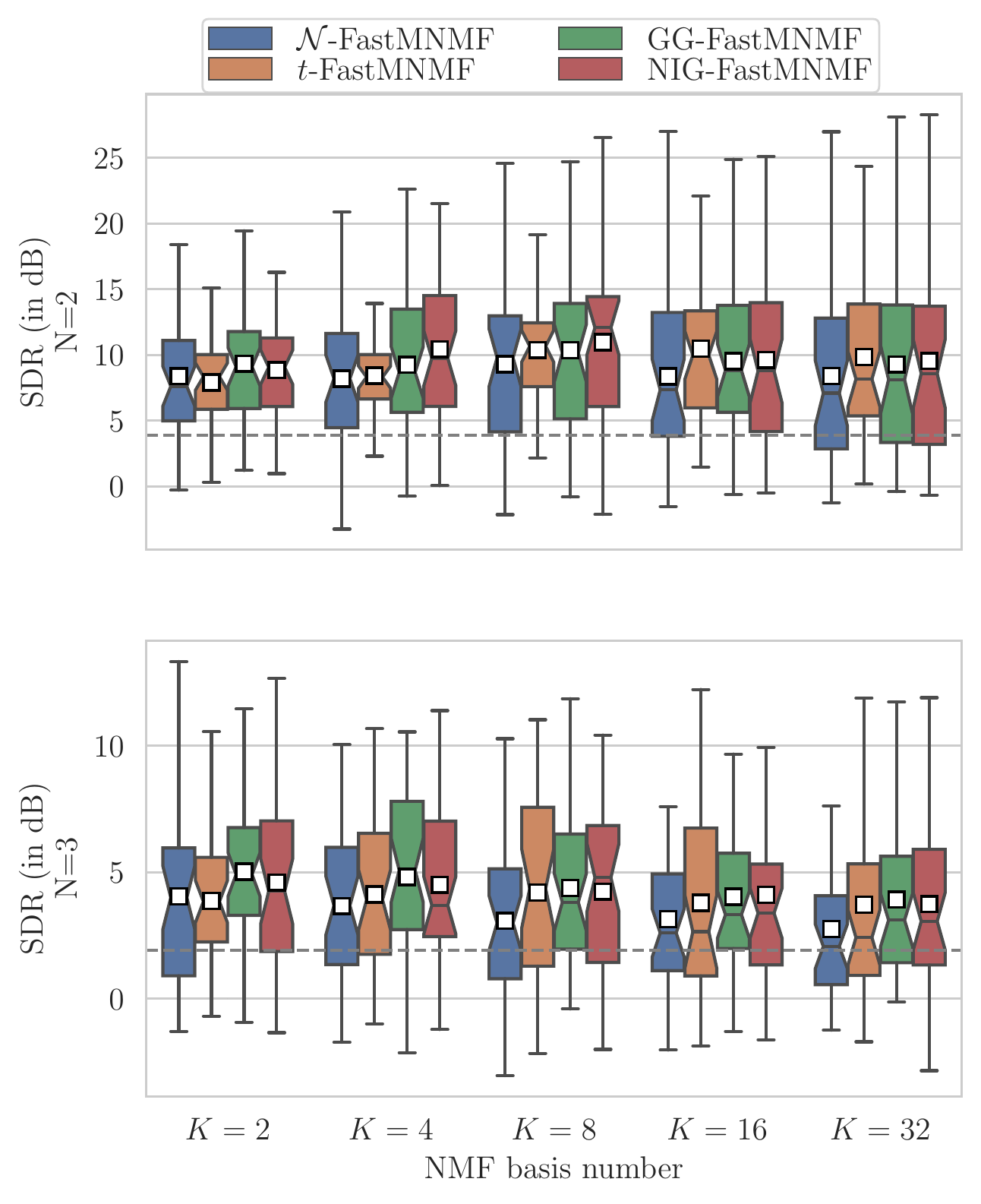}
\caption{
The SDRs obtained by $\mathcal{N}$-, $t$-, GG-, and NIG-FastMNMF with $K\in\{2,4,8,16,32\}$
in speech separation.
White squares and notches indicate the means and the 95\% confidence intervals, respectively. The dashed grey line represents the median SDR results obtained by OverIVA.
}
\label{fig:Init-wsj0}
\end{figure}

The optimal parameters for the speech separation task
    based on the grid search parameter optimization (see Section \ref{sec:settings})
    are shown in Table~\ref{tab:param_sp}.
Fig. \ref{fig:Init-wsj0} shows the SDRs on the validation set
 with $M=8$, $N\in\{2,3\}$, $K\in\{2,4,8,16,32\}$ while Table~\ref{tab:stat_wsj0} reports statistical reference of NIG-FastMNMF with respect to other FastMNMF variants considering a Wilcoxon test.
We discuss the results with $N=2$.
 When $K\in \{4,8\}$,
 NIG-FastMNMF slightly outperformed the other methods in terms of the average and median SDRs with a statistical significance of $ p \approx 0.016$ on average.
 The interquartile range and 95\% confidence interval of NIG-FastMNMF, however, closed one to each other and increased
 as the number of bases $K$ increased. 
  \begin{table}[!h]
\setlength{\tabcolsep}{7pt}
\setlength{\aboverulesep}{1pt}
\setlength{\belowrulesep}{1pt}
\centering
    \caption{Hyperparameters for speech separation}
    \label{tab:param_sp}
    \begin{tabular}{@{}*{4}{c}@{}}
    \toprule
        \multicolumn{4}{c}{FastMNMF variants}\\
        \midrule 
    $\mathcal{N}$ & $t$ & GG & NIG  \\
    n/a & 
    $\nu=100$ &  
    $\beta=1.8$ &  
    $(\rho, \eta)=(15, 1)$    
    \\
    $K=2$ & 
    $K=8$ & 
    $K=2$  &  
    $K=8$   
    \\
    \bottomrule             
    \end{tabular}
\end{table}
We then discuss the results with $N = 3$.
When $K = 2$,
 the interquartile range of GG-FastMNMF
 was smaller than those of the other methods
 with a statistical significance of $ p \approx 0.012$ on average.
GG-, NIG-, and $\mathcal{N}$-FastMNMF with a fewer $K\in\left\{2,4,8\right\}$ 
 yielded better median SDRs,
 whereas $t$-FastMNMF with a larger $K \in\left\{16,32\right\}$ performed better.
Although the median and average SDRs of NIG-FastMNMF with $K=32$ 
 were slightly worse than those of GG-FastMNMF, NIG- and GG-FastMNMF generally tended to perform comparably.
Overall, we found that the best performances of these FastMNMF variants
 were drawn when  $K\in\left\{2,4,8\right\}$.

\subsubsection{Investigation of Performances}
\label{sec:speaker_over}

\begin{table*}
\setlength{\tabcolsep}{7.5pt}
\setlength{\aboverulesep}{1pt}
\setlength{\belowrulesep}{1pt}
\begin{centering}
\caption{SDR, SAR, SIR mean (best is bolded) and standard deviation scores for all settings
in Section \ref{sec:speaker_over}}\label{tab:speaker_over}
\begin{tabular}{@{}lcc*{7}{r}@{}}
\toprule
\multirow{2}{*}[-2pt]{$N$} & \multirow{2}{*}[-2pt]{$M$} & \multirow{2}{*}[-2pt]{score} & \multicolumn{4}{c}{{FastMNMF variants}} \\
\cmidrule(lr){4-7}
  &  &  &
  \multicolumn{1}{c}{$\mathcal{N}$} &
  \multicolumn{1}{c}{$t$} &
    \multicolumn{1}{c}{GG} &
  \multicolumn{1}{c}{NIG}
   \\
\midrule

\multirow{9}{*}[-2pt]{2} & \multirow{3}{*}{2} & SDR &
$2.8~(\pm 3.4)$ & 
$2.8~(\pm 2.9)$ & 
$3.6~(\pm 3.3)$ & 
$\mathbf{3.9~(\pm 3.4)}$ 
\\
& & SAR &
$10.0~(\pm 2.6)$ &
$\mathbf{12.9~(\pm 2.7)}$ &
$11.6~(\pm 2.6)$ &
$11.4~(\pm 2.6)$ 
\\

& & SIR &
$6.5~(\pm 4.1)$ & 
$6.0~(\pm 3.6)$ & 
$7.0~(\pm 4.1)$ & 
$\mathbf{7.3~(\pm 4.3)}$  
\\
\cmidrule(lr){2-7}
& \multirow{3}{*}{5} & SDR &
$7.3~(\pm 5.1)$ & 
$8.0~(\pm 5.2)$ &
$\mathbf{8.7~(\pm 5.6)}$ &
$8.6~(\pm 5.8)$ 
\\

& & SAR &
$14.9~(\pm 4.1)$ & 
$\mathbf{18.0~(\pm 4.4)}$ & 
$17.0~(\pm 4.8)$ & 
$17.2~(\pm 5.0)$ 
\\

& & SIR &
$12.0~(\pm 6.0)$ & 
$12.3~(\pm 6.3)$ & 
$13.4~(\pm 6.8)$ & 
$\mathbf{13.4~(\pm 6.7)}$  
\\

\cmidrule(lr){2-7}
& \multirow{3}{*}{8} & SDR &
$7.7~(\pm 5.1)$ & 
$8.3~(\pm 4.9)$ & 
$8.9~(\pm 5.8)$ & 
$\mathbf{9.4~(\pm 5.6)}$  
\\

& & SAR &
$16.7~(\pm 4.3)$ & 
$\mathbf{19.2~(\pm 4.2)}$ & 
$18.7~(\pm 5.1)$ & 
$19.0~(\pm 4.8)$ 
\\

& & SIR &
$12.7~(\pm 6.7)$  &
$12.8~(\pm 6.6)$ & 
$14.0~(\pm 7.6)$ & 
$\mathbf{14.3~(\pm 7.3)}$ 
\\
\midrule
\multirow{9}{*}[-2pt]{3} & \multirow{3}{*}{3} & SDR &
$1.2~(\pm 2.0)$ & 
$1.0~(\pm 2.1)$ & 
$1.3~(\pm 2.1)$ & 
$\mathbf{1.5~(\pm 2.3)}$ 
\\

& & SAR &
$8.6~(\pm 2.0)$ & 
$10.0~(\pm 1.5)$ & 
$\mathbf{11.3~(\pm 1.7)}$ & 
$9.9~(\pm 1.6)$ 
\\

& & SIR &
$3.7~(\pm 3.0)$ & 
$3.0~(\pm 2.8)$ & 
$3.4~(\pm 2.9)$ & 
$\mathbf{3.7~(\pm 2.4)}$ 
\\
\cmidrule(lr){2-7}
& \multirow{3}{*}{5} & SDR &
$2.8~(\pm 3.2)$ & 
$3.1~(\pm 3.4)$ & 
$3.3~(\pm 3.1)$ & 
$\mathbf{3.5~(\pm 3.2)}$ 
\\

& & SAR &
$12.9~(\pm 2.7)$ &
$12.8~(\pm 2.4)$ &
$\mathbf{14.1~(\pm 2.8)}$ &
$11.1~(\pm 2.9)$
\\
& & SIR &
$6.1~(\pm 4.3)$ & 
$5.9~(\pm 4.2)$ & 
$\mathbf{6.5~(\pm 4.4)}$ & 
$6.2~(\pm 4.3)$ 
\\

\cmidrule(lr){2-7}
& \multirow{3}{*}{8} & SDR &
$4.5~(\pm 3.8)$ & 
$4.5~(\pm 3.6)$ & 
$5.0~(\pm 3.8)$ & 
$\mathbf{5.1~(\pm 3.7)}$ 
\\

& & SAR &
$13.8~(\pm 3.6)$ & 
$\mathbf{16.0~(\pm 3.2)}$ & 
$15.6~(\pm 3.4)$ & 
$15.7~(\pm 3.4)$  
\\
& & SIR &
$8.3~(\pm 5.0)$ & 
$7.6~(\pm 4.9)$ & 
$\mathbf{8.6~(\pm 5.2)}$ & 
$8.5~(\pm 5.1)$ 
\\

\bottomrule
\end{tabular}
\par\end{centering}
\end{table*}

Table~\ref{tab:speaker_over} shows the SDRs, SARs, and SIRs on the test set
 obtained by the four methods with the optimized hyperparameters.
Overall, 
 NIG-FastMNMF attained the best SDRs and SIRs,
 whereas $t$-FastMNMF attained the best SARs. 
The numerically-unstable computation of the modified Bessel function 
 may have affected the SAR of NIG-FastMNMF.
For $N=2$,
 the SDR improvement from $M=5$ to $M=8$
 was small for $t$- and $\mathcal{N}$-FastMNMF,
 whereas that was more significant for GG- and NIG-FastMNMF.

Considering the overall results from investigation in Section \ref{sec:speech_enhancement} and \ref{sec:speech_separation},
 the proposed NIG-FastMNMF 
 can be claimed as being the most reasonable choice with an adequate set of hyperparameters for speech separation
 as well as speech enhancement.

\begin{table}[!tp]
\setlength{\tabcolsep}{7.5pt}
\setlength{\aboverulesep}{1pt}
\setlength{\belowrulesep}{1pt}
\begin{centering}
\caption{ Statistical significance (  "***" denotes high ($p< 0.001$), "**" good ($p<0.01$), "*" marginal ($p<0.05$) and "n.s." non significant ($p>=0.05$) p-value)
 for a non-parametric Wilcoxon tested on the NIG-FastMNMF SDR scores obtained in Section \ref{sec:speaker_over}}\label{tab:stat_wsj0}
\begin{tabular}{@{}lcc*{7}{r}@{}}
\toprule
\multirow{2}{*}[-2pt]{$N$} & \multirow{2}{*}[-2pt]{$K$}  & \multicolumn{3}{c}{{FastMNMF variants}}
\\
\cmidrule(lr){3-5}
  &  &
  \multicolumn{1}{c}{$\mathcal{N}$} &
  \multicolumn{1}{c}{$t$} &
    \multicolumn{1}{c}{GG}
   \\
\midrule

\multirow{5}{*}[-2pt]{2} & 2 &
* &
***  &
n.s.
\\
&  4 &
** &
*  &
**
\\

&  8 &
*** &
*  &
**
\\
&  16 &
*** &
**  &
n.s.
\\

&  32 &
*** &
n.s.  &
*
\\
\midrule\\ 
\multirow{5}{*}{3} & 2 &
*** &
**  &
*
\\
&  4 &
** &
*  &
n.s.
\\

&  8 &
*** &
n.s  &
*
\\
&  16 &
*** &
n.s.  &
n.s.
\\

&  32 &
** &
**  &
*
\\

\bottomrule
\end{tabular}
\par\end{centering}
\end{table}


\section{Conclusion}
\label{sec:conclusion}

This paper has described GSM-FastMNMF, 
 a robust generalization of Gaussian FastMNMF 
 ($\mathcal{N}$-FastMNMF),
 that incorporates a general expression of heavy-tailed probability distributions
 called a Gaussian scale mixture (GSM) 
 into the jointly-diagonalizable spatial model FastMNMF.
We have developed a multiplicative update variational expectation-maximization (MU-VEM) algorithm 
 for GSM-FastMNMF.
As an instance of GSM-FastMNMF,
 we have derived generalized hyperbolic FastMNMF (GH-FastMNMF),
 which encompasses not only $\mathcal{N}$-FastMNMF and Student's $t$ FastMNMF ($t$-FastMNMF)
 but also a new variant called normal-inverse Gaussian FastMNMF (NIG-FastMNMF).
We showed that leptokurtic generalized Gaussian FastMNMF (GG-FastMNMF),
 which does not belong to GH-FastMNMF,
 can also be instantiated from GSM-FastMNMF.
The speech enhancement and separation results
 revealed the experimental advantages of NIG-FastMNMF in most conditions.

Considering the recent advance of deep learning techniques, 
 one important future direction is to use a normalizing flow~\cite{rezende2015variational}
 for formulating an adaptive time-varying spatial model
 as proposed in \cite{nugraha2020flow}.
Another complementary direction is to use a deep generative model of speech 
 for improving the expression capability of the source model 
 as proposed in \cite{fontaine2019cauchy, leglaive2019speech}. 
From the laborious grid study of this paper, a next step will be also to estimate the tail-index parameters of a given mixture $\mbX$ as in \cite{snoussi2006bayesian}.  

The proposed general formalism of GSM-FastMNMF
 could be extended for other scale mixture models 
 such as the generalized Gaussian scale mixture \cite{gupta2018generalized}. 
 

\appendix


\section*{Probability Density Functions of\\Gaussian Scale Mixture Variables}

Let $\mbx \in \mathbb{C}^M$ be a $M$-dimensional complex random vector
 following a zero-mean elliptically-contoured multivariate 
 complex Gaussian scale mixture (GSM) 
 with a positive semidefinite scale matrix $\mbSi \succeq \bm{0}$.
Concrete examples are described below:
\begin{itemize}
    \item A centralized Gaussian distribution is denoted by $\mbx \sim \ComplexGaussian{\mbSi}$ 
    and the PDF of $\mbx$ is given by
    \begin{align}
        p(\mbx) 
        = 
        \frac{1}{\pi^{M}|\mbSi|}
        \exp\!\left(-\mbx^{\Hr}\mbSi^{-1}\mbx\right). 
        \label{eq:pdf-N}
    \end{align}
    \item A Student's $t$ distribution with a degree of freedom $\nu > 0$
    is denoted by $\mbx \sim \ComplexStudent{\nu}{\mbSi}$ 
    and the PDF of $\mbx$ is given by
    \begin{align}
        p(\mbx)
        =
        \frac{2^{M}\Gamma\!\left(\frac{2M+\nu}{2}\right)}{\left(\pi\nu\right)^{M} \Gamma\!\left(\frac{\nu}{2}\right) |\mbSi|}
        \left(1 + \frac{2}{\nu}\mbx^{\Hr}\mbSi^{-1}\mbx\right)^{-\frac{2M+\nu}{2}}.
        \label{eq:pdf-T}
    \end{align}
    \item A generalized Gaussian (GG) distribution with a shape parameter $\beta > 0$
    is denoted by $\mbx \sim \ComplexGG{\beta}{\mbSi}$ 
    and the PDF of $\mbx$ is given by
    \begin{align}
        p(\mbx) 
        =
        \frac{\frac{\beta}{2}\Gamma\!\left(M\right)}{2^{\frac{2M}{\beta}}\pi^{M} \Gamma\!\left(\frac{2M}{\beta}\right)|\mbSi|}
        \exp\!\left(-\left(\mbx^{\Hr}\mbSi^{-1}\mbx\right)^{\frac{\beta}{2}}\right). \label{eq:pdf-GG}
    \end{align}
    \item A generalized hyperbolic (GH) distribution 
    with a shape parameter $\gamma \in \mathbb{R}$,
    a concentration parameter $\rho > 0$,
    and a scaling parameter $\eta > 0$
    is denoted by $\mbx \sim \ComplexGH{\gamma,\rho,\eta}{\mbSi}$ 
    and the PDF of $\mbx$ is given by
    \begin{align}
        p(\mbx) 
         =&
        \frac{1}{(\pi \eta)^M \mathcal{K}_{\gamma}(\rho)|\mbSi|}
        \left(1 + \frac{2}{\rho\eta}\mbx^{\mathrm{H}}\mbSi^{-1}\mbx\right)^{\frac{\gamma-M}{2}}
        \nonumber\\
        &\mathcal{K}_{\gamma-M}\!\left(\rho\eta^{-1} \sqrt{\rho\eta + 2\mbx^{\mathrm{H}}\mbSi^{-1}\mbx}\right). \label{eq:pdf-GH}
    \end{align}
\end{itemize}

\section*{Probability Density Functions of\\Impulse Variables}

Let $x$ be a nonnegative random variable.
Concrete examples are described below:
\begin{itemize}

    \item An inverse gamma (IG) distribution with a shape parameter $\alpha > 0$ and a scale parameter $\sigma > 0$
    is denoted by $x \sim \InverseGamma{\alpha}{\sigma}$
    and the PDF of $x$ is given by
    \begin{align}
        p(x)
        =
        \frac{\sigma^\alpha}{\Gamma(\alpha)}
        x^{- \alpha - 1} e^{- \sigma x^{-1}}.\label{eq:pdf_IG}
    \end{align}

    \item A generalized inverse Gaussian (GIG) distribution
    with a shape parameter $\gamma \in \mathbb{R}$,
    a concentration parameter $\rho > 0$,
    and a scaling parameter $\eta > 0$
    is denoted by $x \sim \GeneralizedInverseGaussian{\gamma}{\rho}{\eta}$
    and the PDF of $x$ is given by
    \begin{align}
        p(x)
        =
        \frac{1}{{2 \eta^{\gamma} \mathcal{K}_{\gamma}(\rho)}}
        x^{\gamma-1}
        e^{-\frac{\rho}{2}\left(\eta^{-1} x + \eta x^{-1}\right)}.
        \label{eq:pdf_GIG}
    \end{align}
\end{itemize}

\section*{Proof of Eq. \eqref{eq:e_log_p_zft}}

Let $\mbz_{ft} \in \mathbb{C}^M$ be an $M$-dimensional complex random vector 
 drawn from a Gaussian scale mixture (GSM)  
 as described in Section~\ref{sec:GSM}. 
The gradient of $p(\mbz_{ft})$ is given by \cite{palmer2007modeling}
\begin{align}
\frac{d}{d\mbz_{ft}^{\Hr}} p(\mbz_{ft}) 
&=
\frac{d}{d\mbz_{ft}^{\Hr}}\int p(\mbz_{ft}\mid\phi_{ft})p(\phi_{ft})d\phi_{ft}
\nonumber\\
&=
\int
p(\phi_{ft})
\frac{d}{d\mbz_{ft}^{\Hr}}
p(\mbz_{ft}\mid\phi_{ft}) d\phi_{ft}.
\label{eq:dif-Gau}
\end{align}
Because $p(\mbz_{ft}\mid \phi_{ft})$ is an isotropic complex Gaussian distribution, 
 its derivative is given by
\begin{align}
\frac{d}{d\mbz_{ft}^{\Hr}} p(\mbz_{ft}\mid\phi_{ft})
=
- 2\tilde{\mbY}_{ft}^{-1}\mbz_{ft} \phi_{ft}^{-1} p(\mbz_{ft}\mid\phi_{ft}),
\label{eq:dif-condGau}
\end{align}
where $\tilde{\mbY}_{ft}$ is given in Eq.~\eqref{eq:p_z_ft}.
Substituting Eq.~\eqref{eq:dif-condGau} into Eq.~\eqref{eq:dif-Gau}, we obtain
\begin{align}
\frac{d}{d\mbz_{ft}^{\Hr}} p(\mbz_{ft}) 
 & =-2\tilde{\mbY}_{ft}^{-1}\mbz_{ft}\int\phi_{ft}^{-1}p(\mbz_{ft}, \phi_{ft})d\phi_{ft}
 \nonumber\\
 & =-2\tilde{\mbY}_{ft}^{-1}\mbz_{ft}p(\mbz_{ft})\int\phi_{ft}^{-1}p(\phi_{ft}\mid\mbz_{ft})d\phi_{ft}
 \nonumber\\
 & =-2\tilde{\mbY}_{ft}^{-1}\mbz_{ft}p(\mbz_{ft})\mathbb{E}_{p(\phi_{ft}\mid\mbz_{ft})}\big[\phi_{ft}^{-1}\big].
 \label{eq:dif-Gau2}
\end{align}
Using Eq.~\eqref{eq:dif-Gau2}, we have
\begin{align}
\frac{d}{d\mbz_{ft}^{\Hr}} \log p(\mbz_{ft}) 
&= 
p(\mbz_{ft})^{-1}
\frac{d}{d\mbz_{ft}^{\Hr}} p(\mbz_{ft}),
\nonumber\\
&=
-2\tilde{\mbY}_{ft}^{-1}\mbz_{ft} 
\mathbb{E}_{p(\phi_{ft}\mid\mbz_{ft})}\big[\phi_{ft}^{-1}\big].
\end{align}
This proves Eq.~\eqref{eq:e_log_p_zft}.

\section*{Proof of Eq. \eqref{eq:log_likelihood_jdsm_gh}}

In the same way 
 as a multivariate real generalized hyperbolic (GH) distribution \cite{hammerstein2010generalized} ,
 an isotropic multivariate complex GH distribution $p(\mbx)$ of dimension $M$
is given by perturbing the scale of 
 an isotropic multivariate complex Gaussian distribution $p(\mbx | \phi)$
 with a generalized inverse Gaussian (GIG) distribution $p(\phi)$ as follows
 :
 
\begin{align}
p(\mbx)
&=
\int_{0}^{\infty}p(\mbx\mid\phi)p(\phi)d\phi,
\label{eq:marg_dist}
\\
p(\mbx\mid\phi)
&=
\frac{1}{\pi^{M}|\phi\mbSi|}
e^{-\mbx^{\Hr}\left(\phi\mbSi\right)^{-1}\mbx},
\label{eq:cond_Gauss}
\\
p(\phi)
&=
\frac{1}{2\eta^{\gamma}\mathcal{K}_{\gamma}(\rho)}\phi^{\gamma-1}e^{-\frac{\rho}{2}\left(\eta^{-1}\phi+\eta\phi^{-1}\right)},
\label{eq:GIG}
\end{align}
where $\mbSi \succeq \bm{0}$ is a positive semidefinite matrix of dimension $M$
 and $\gamma \in \mathbb{R}$,
 $\rho > 0$,
$\eta > 0$ are the GIG parameters.
 Eq. \eqref{eq:marg_dist} is computed as follows:
\begin{align}
p(\mbx) & =C_{\gamma,\rho,\eta,\mbSi}\int_{0}^{\infty}\phi^{\gamma-M-1}e^{-\frac{1}{2}\left(\frac{1}{\phi}\left(2\mbx^{\Hr}\mbSi^{-1}\mbx+\rho\eta\right)+\rho\eta^{-1}\phi\right)}d\phi\nonumber\\
 & =C_{\gamma,\rho,\eta,\mbSi}\left(\frac{2\mbx^{\Hr}\mbSi^{-1}\mbx+\rho\eta}{\rho\eta^{-1}}\right)^{\frac{\gamma-M}{2}}\nonumber\\
 & \quad\int\psi^{\gamma-M-1}e^{-\frac{1}{2}\left(\left(\text{\text{\ensuremath{\frac{1}{\phi}}+\ensuremath{\phi}}}\right)\sqrt{\rho^{2}+2\rho\eta^{-1}\mbx^{\Hr}\mbSi^{-1}\mbx}\right)}d\psi\nonumber\\
 & =2C_{\gamma,\rho,\eta,\mbSi}\left(\frac{2\mbx^{\Hr}\mbSi^{-1}\mbx+\rho\eta}{\rho\eta^{-1}}\right)^{\frac{\gamma-M}{2}}\nonumber\\
 & \quad\quad\mathcal{K}_{\gamma-M}\left(\sqrt{\rho^{2}+2\rho\eta^{-1}\mbx^{\Hr}\mbSi^{-1}\mbx}\right)
    \label{eq:const_pdf_GH}
\end{align}
where $C_{\gamma,\rho,\eta,\mbSi} = \frac{1}{2\eta^{\gamma}\mathcal{K}_{\gamma}(\rho)\pi^{M}\left|\mbSi\right|}$ and the substitution $\psi = \sqrt{\frac{\rho\eta^{-1}}{2\mbx^{\Hr}\mbSi^{-1}\mbx+\rho\eta}}\phi$ occurs on the second equality. The integral in the second equality is finally calculated thanks to the relation\cite{hammerstein2010generalized} as follows:
\begin{align}
\forall \theta > 0,~
    \mathcal{K}_{k}\left(\theta\right)
    =
    \frac{1}{2}\int_{0}^{\infty}q^{k-1}e^{-\frac{1}{2}\left(\frac{1}{q}+q\right)\theta}dq.
\end{align}
Eq.~\eqref{eq:log_likelihood_jdsm_gh} can be simply proved by introducing the FastMNMF model and their variables $\tilde{z}_{ftm}$ and $\tilde{y}_{ftm}$ defined in Eqs.~\eqref{eq:xti_ftm} and \eqref{eq:yti_ftm}, respectively.

\bibliographystyle{IEEEtran}
\bibliography{IEEEabrv, biblio}

\begin{IEEEbiography}[{\includegraphics[width=1in,height=1.25in,clip,keepaspectratio]{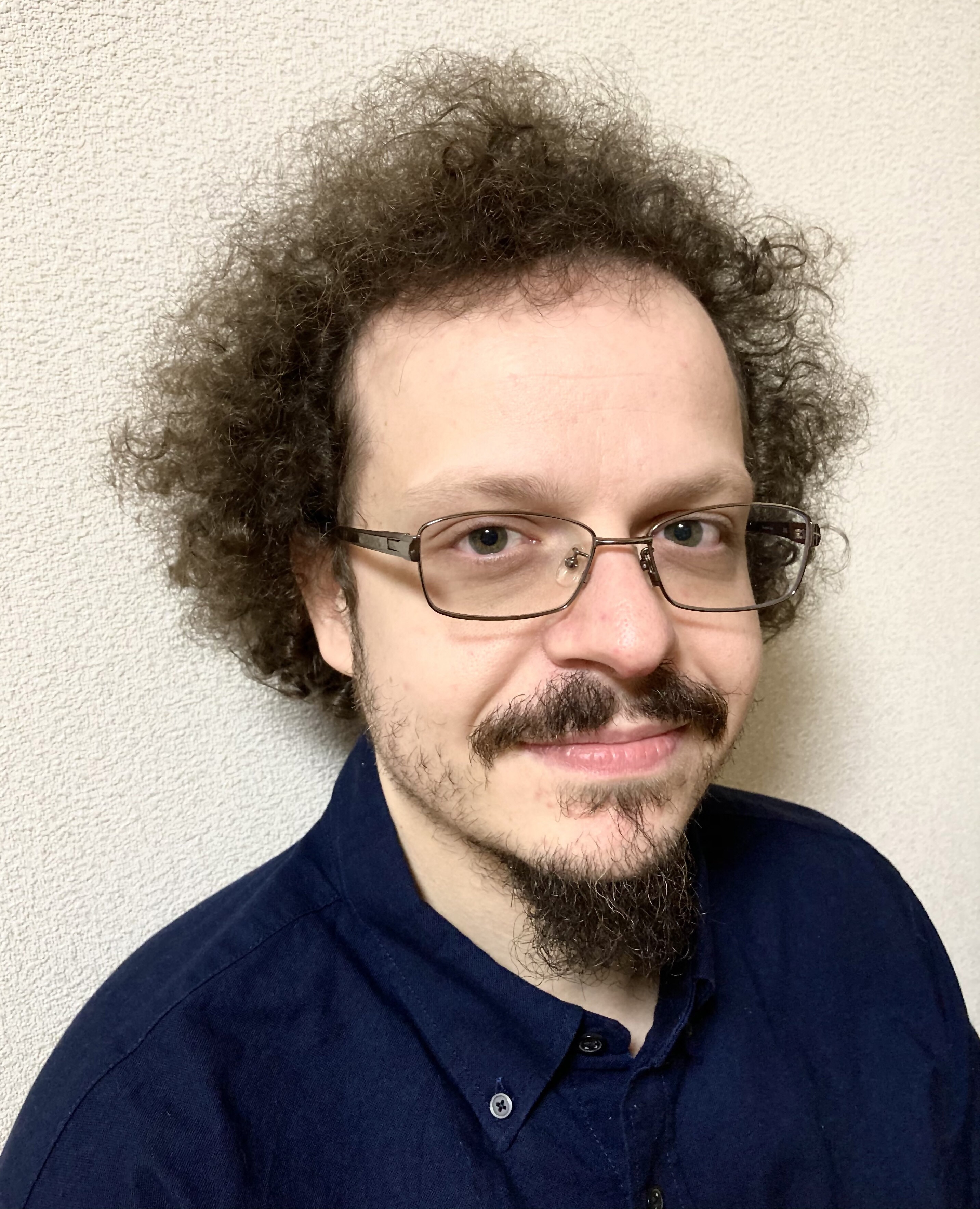}}]{Mathieu Fontaine}
received the M.S. degree in Applied \& Fundamentals Mathematics from Universit\'{e} de Poitiers, Poitiers, France, in 2015 and the Ph.D. degree in informatics from Universit\'{e} de Lorraine and Inria Nancy Grand-Est, France, in 2018 and was a Postdoctoral Researcher at the Center for Advanced Intelligence Project (AIP), RIKEN, Japan.
He is currently an assistant professor at LTCI, Télécom Paris, Palaiseau, France.
He is also a visiting researcher at the Advanced Intelligence Project (AIP), RIKEN, Tokyo, Japan.
His research interests include machine listening topics such as audio source separation, sound event detection and speaker diarization using microphone array.
\end{IEEEbiography}

\begin{IEEEbiography}[{\includegraphics[width=1in,height=1.25in,clip,keepaspectratio]{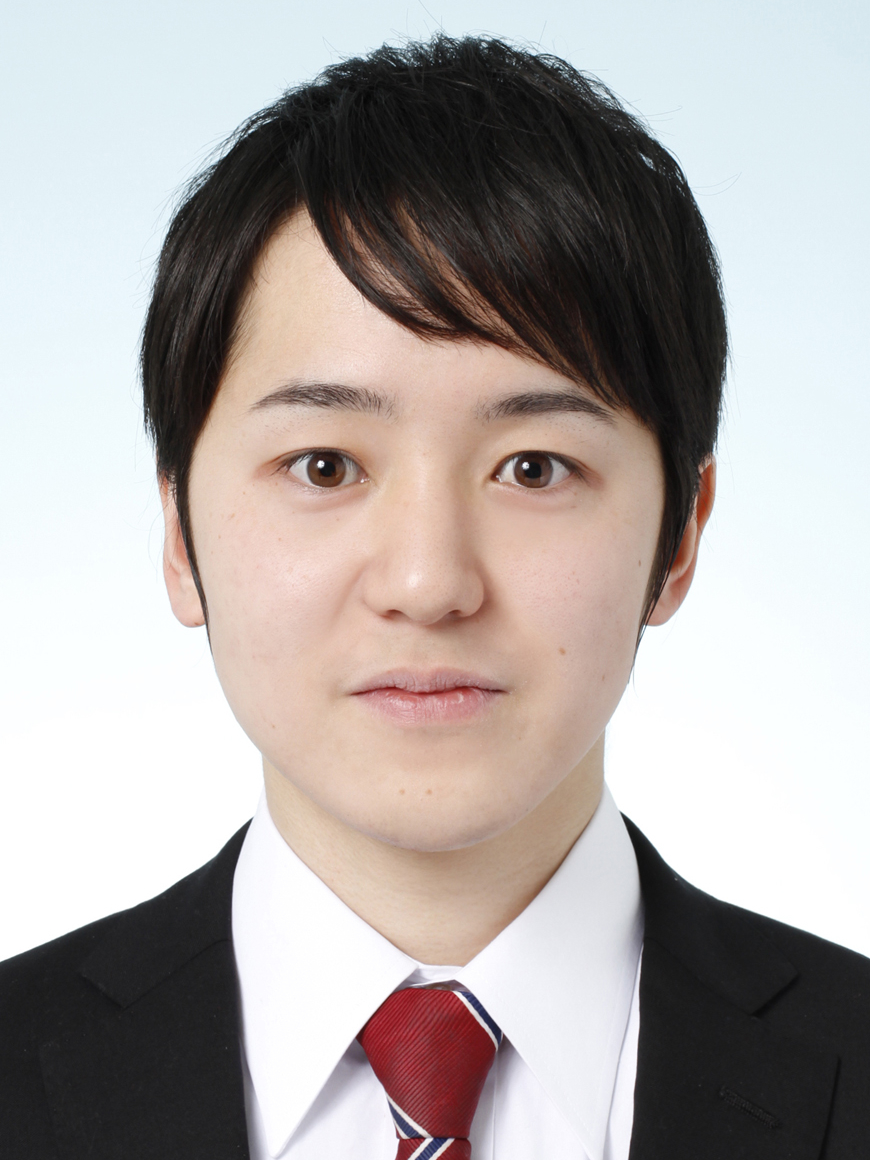}}]{Kouhei Sekiguchi}
received the B.E. and M.S. degrees from Kyoto University, Kyoto, Japan, in 2015 and 2017, respectively and the Ph.D. degree from Kyoto University in 2021.
He is currently a Postdoctoral researcher at the Center for Advanced Intelligence Project (AIP), RIKEN, Japan.
His research interests include microphone array signal processing and machine learning.
He is a member of IEEE and IPSJ.
\end{IEEEbiography}

\begin{IEEEbiography}[{\includegraphics[width=1in,height=1.25in,clip,keepaspectratio]{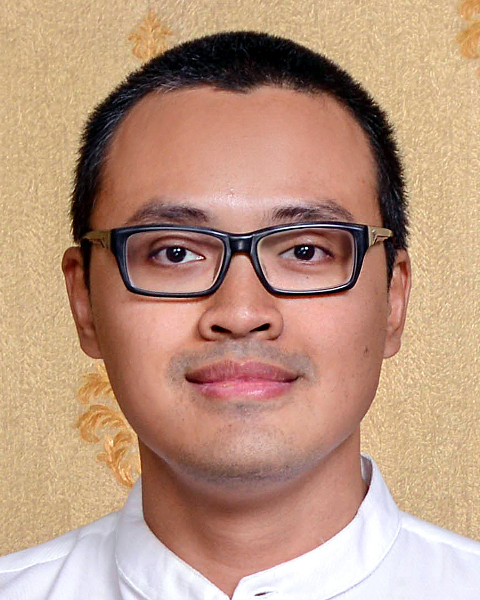}}]{Aditya Arie Nugraha}
received the B.S. and M.S. degrees in electrical engineering from Institut Teknologi Bandung, Indonesia, in 2008 and 2011, respectively, 
the M.E. degree in computer science and engineering from Toyohashi University of Technology, Japan, in 2013, 
and the Ph.D. degree in informatics from Universit\'{e} de Lorraine and Inria Nancy--Grand-Est, France, in 2017. 
He is currently a Research Scientist in the Sound Scene Understanding Team at the Center for Advanced Intelligence Project (AIP), RIKEN, Japan.
His research interests include audio-visual signal processing and machine learning.
\end{IEEEbiography}

\begin{IEEEbiography}[{\includegraphics[width=1in,height=1.25in,clip,keepaspectratio]{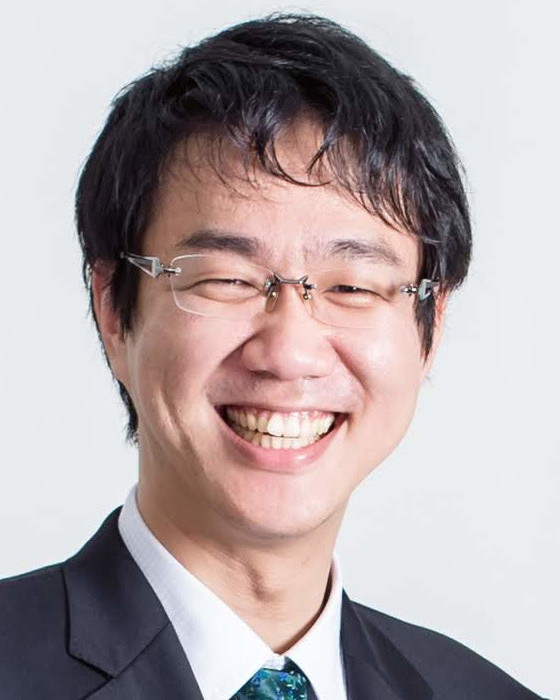}}]{Yoshiaki Bando}
received the M.S. and Ph.D. degrees in informatics from Kyoto University, Kyoto, Japan, in 2015 and 2018, respectively.
He is currently a senior researcher at Artificial Intelligence Research Center (AIRC), 
National Institute of Advanced Industrial Science and Technology (AIST), Tokyo, Japan.
He is also a visiting researcher at the Advanced Intelligence Project (AIP), RIKEN, Tokyo, Japan.
His research interests include microphone array signal processing, deep Bayesian learning, and robot audition.
He is a member of IEEE, RSJ, and IPSJ.
\end{IEEEbiography}

\begin{IEEEbiography}[{\includegraphics[width=1in,height=1.25in,clip,keepaspectratio]{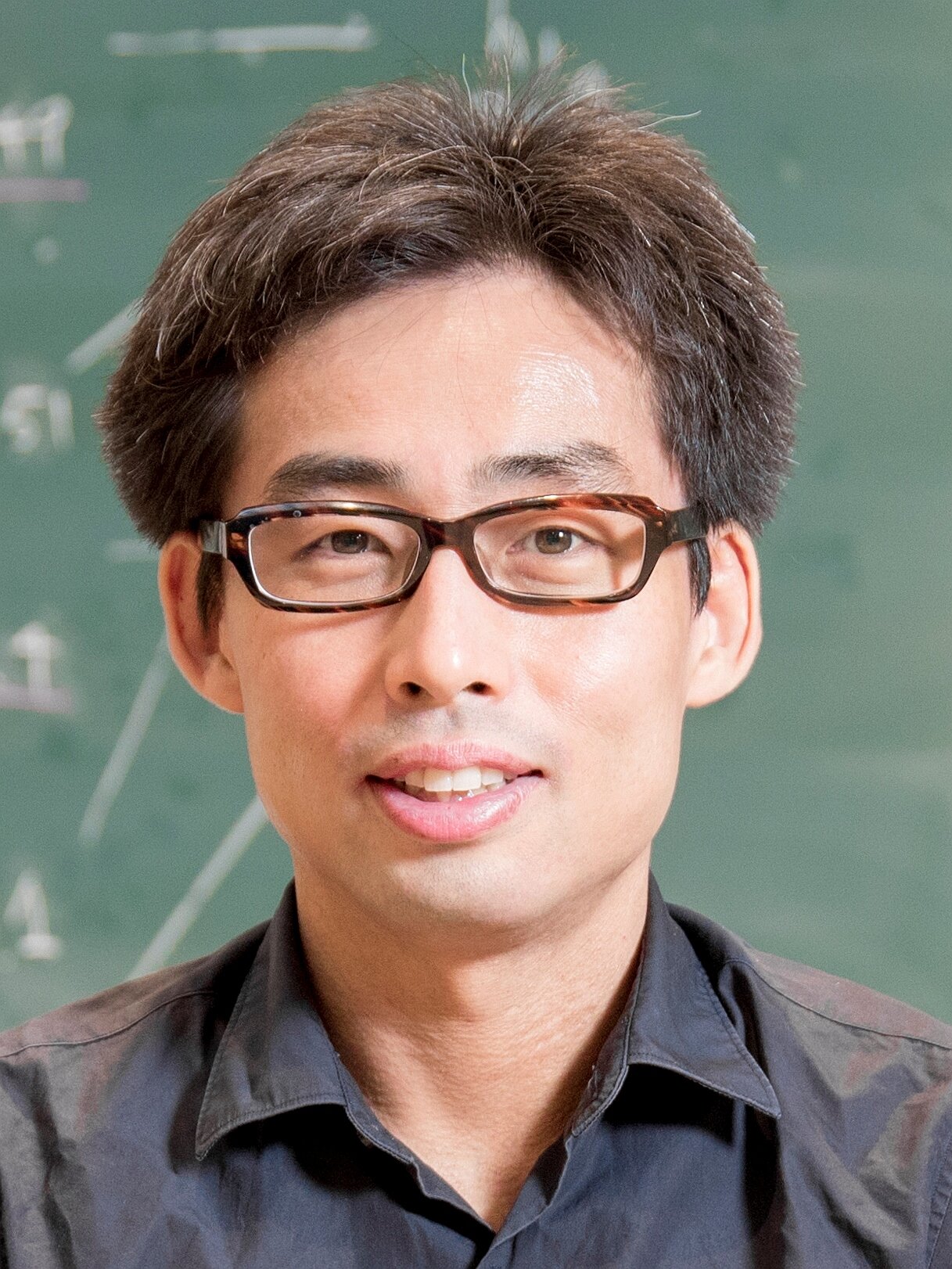}}]{Kazuyoshi Yoshii}
received the M.S. and Ph.D. degrees in informatics from Kyoto University, Kyoto, Japan, in 2005 and 2008, respectively.
He is an Associate Professor at the Graduate School of Informatics, Kyoto University,
and concurrently the Leader of the Sound Scene Understanding Team,
Center for Advanced Intelligence Project (AIP), RIKEN, Tokyo, Japan.
His research interests include music informatics,
audio signal processing, and statistical machine learning.
\end{IEEEbiography}

\end{document}